\documentclass[aps,prl,reprint,superscriptaddress]{revtex4-2}

\usepackage[english]{babel}
\usepackage{graphics}
\usepackage{array}
\usepackage{graphicx}
\usepackage{todonotes}
\usepackage{amsfonts,amssymb,amsmath}
\usepackage{color}
\usepackage{threeparttable}
\usepackage{physics}
\usepackage{comment}
\usepackage[separate-uncertainty,per-mode=fraction]{siunitx}

\definecolor{darkblue}{rgb}{0.0,0.0,0.4}
\definecolor{darkgreen}{rgb}{0.0,0.4,0.0}
\definecolor{darkred}{rgb}{0.6,0.0,0.0}
\usepackage[bookmarksopen]{hyperref}
\hypersetup{colorlinks,linkcolor=darkblue,citecolor=darkblue,urlcolor=darkblue}

\newcommand{\gs}{$\mathrm{X}^2\Sigma^+$}
\newcommand{\exs}{$\mathrm{A}^2\Pi_{1/2}$}
\newcommand{\baff}[1]{${}^{#1}\mathrm{BaF}$}

\newcolumntype{M}[1]{>{\centering\arraybackslash}m{#1}}

\makeatletter
\newcommand\thefontsize[1]{{#1 The current font size is: \f@size pt\par}}
\makeatother

\begin{document}

\title{Laser cooled $^{137}$BaF molecules\\ for measuring nuclear-spin-dependent parity violation
}

\author{Felix Kogel}
\affiliation{5. Physikalisches  Institut  and  Center  for  Integrated  Quantum  Science  and  Technology, Universit\"at  Stuttgart,  Pfaffenwaldring  57,  70569  Stuttgart,  Germany}

\author{Tatsam Garg}
\affiliation{5. Physikalisches  Institut  and Center  for  Integrated  Quantum  Science  and  Technology, Universit\"at  Stuttgart,  Pfaffenwaldring  57,  70569  Stuttgart,  Germany}

\affiliation{Vienna Center for Quantum Science and Technology, Atominstitut, TU Wien,  Stadionallee 2,  A-1020 Vienna,  Austria}

\author{Marian Rockenh\"auser}
\affiliation{5. Physikalisches  Institut  and Center  for  Integrated  Quantum  Science  and  Technology, Universit\"at  Stuttgart,  Pfaffenwaldring  57,  70569  Stuttgart,  Germany}

\affiliation{Vienna Center for Quantum Science and Technology, Atominstitut, TU Wien,  Stadionallee 2,  A-1020 Vienna,  Austria}

\author{Tim Langen}
\email{tim.langen@tuwien.ac.at}

\affiliation{5. Physikalisches  Institut  and  Center  for  Integrated  Quantum  Science  and  Technology, Universit\"at  Stuttgart,  Pfaffenwaldring  57,  70569  Stuttgart,  Germany}

\affiliation{Vienna Center for Quantum Science and Technology, Atominstitut, TU Wien,  Stadionallee 2,  A-1020 Vienna,  Austria}

\begin{abstract}
We demonstrate optical cycling and transverse laser cooling of a beam of fermionic \baff{137} molecules. Their high masses and nuclear spins make these molecules sensitive probes for parity violation and properties of the weak interaction. However, the nuclear spins also lead to a quasi-closed cycling transition currently involving up to $112$ levels, which significantly exceeds the complexity in other laser-cooled molecules. Optical cycling and cooling are facilitated through carefully designed optical spectra tailored to this molecular structure. Our results pave the way for efficient state preparation, detection, and cooling in precision measurements using this species and other similar species.
\end{abstract}

\maketitle

\textit{Introduction---}The sensing capabilities of molecules have recently been established in a series of landmark experiments, which have advanced the search for a permanent electric dipole moment of the electron~\cite{Hudson2011,ACME2018,Roussy2023}, constrained the variation of fundamental constants~\cite{Shelkovnikov2008,Chin2009,Truppe2017,Kobayashi2019} and provided precise tests of quantum electrodynamics~\cite{Alighanbari2020,Germann2021,Doran2024}. These and related experiments have enabled stringent tests for physics beyond the Standard Model (SM)~\cite{DeMille2017,Safronova2018,Demille2024}. The application of laser cooling to more molecular species, which so far---with few exceptions~\cite{Lim2018,Alauze2021,McNally2020,Rockenhaeuser2024,Lasner2024}---has been limited to comparably light and simple diatomic and polyatomic species, is expected to further increase the sensitivity of such tests in the future~\cite{Kozyryev2017,Fitch2021methods,Anderegg2024,Fitch2021}.

Measurements of parity violation, particularly nuclear-spin-dependent parity violation, in atoms and molecules can extend these tests to the electroweak sector of the SM without interference from significant electromagnetic backgrounds~\cite{Bouchiat1997}. This is expected to enable the precise determination of several coupling parameters that remain poorly constrained by accelerator-based experiments~\cite{Safronova2018,Wang2014,Altuntas2018}. To date, parity violation has been observed in several heavy atomic species~\cite{Bouchiat1997,Antypas2019}, but only a single experiment using ${}^{133}$Cs has observed nuclear-spin-dependent effects~\cite{Wood1997}, finding notably different values for the corresponding SM parameters than other nuclear parity violation experiments~\cite{Haxton2002,Johnson2003}. 

Recently, it has been experimentally demonstrated that barium monofluoride (BaF) molecules can enhance nuclear-spin-dependent parity violation effects by more than $10$ orders of magnitude~\cite{Kozlov1995,DeMille2008,Altuntas2018,AltuntasPRA}. However, due to the low abundance of the relevant odd isotopologues of BaF, and populations distributed over a large number of hyperfine levels in these isotopologues, the corresponding experiments face significant challenges and have so far remained limited to the most abundant even isotopologue \baff{138}. In this isotopologue, the effects of parity violation are vanishingly small due to the absence of a nuclear spin in the even barium atom and the negligible valence electron density at the fluorine nucleus~\cite{DeMille2008}.

Here, we demonstrate optical cycling and transverse laser cooling of odd \baff{137}, which is expected to show significant nuclear-spin-dependent parity violation connected to the finite spin of the $^{137}$Ba nucleus~\cite{DeMille2008}. The principles described in the following offer a promising solution to overcome the limitations that have so far prevented the detection of parity violation in this and similar molecules, and could increase the sensitivity of experiments to the level needed to search for physics beyond the SM in the weak sector~\cite{Dzuba2017,Gaul2020}. 

\textit{Measurement strategy---}The enhancement of nuclear-spin-dependent parity violation in molecules relies on the inherent presence of closely spaced rotational states with opposite parity. These states can be tuned to near degeneracy using moderate magnetic fields, which greatly enhances the parity-violation-induced mixing of these states~\cite{Flambaum1985,DeMille2008}. In practice, the sensitivity for detecting this mixing is limited by the achievable spatial extent and homogeneity of the magnetic field. In conventional beam experiments, molecules are required to traverse an interaction region of considerable length, often extending several meters, prior to the analysis of their internal state~\cite{Altuntas2018}. The first motivation motivation for laser cooling is thus to maximize the flux of molecules passing through this interaction region by reducing their transverse divergence, which can lead to an improvement of several orders of magnitude in the detectable molecules after the interaction region~\cite{Kogel2021,Alauze2021}. Second, and even more desirable for the future, laser slowing and trapping could confine the molecules in a small volume for extended periods of time, eliminating many of the stringent requirements for magnetic field homogeneity and significantly increasing interaction times~\cite{Norrgard2019,Fitch2021methods,Karthein2024}.

\begin{figure}[tb]
    \centering
    \includegraphics[width=0.49\textwidth]{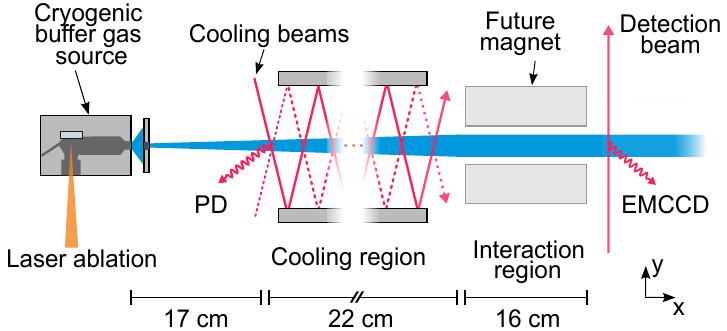}
    \caption{Experimental setup. An isotopically mixed molecular beam of BaF molecules is created by laser ablation in a cryogenic buffer gas source. Following collimation and expansion, the molecules traverse a cooling region, where \baff{137} molecules interact with retroreflected laser beams --- a single beam (dashed line) for optical cycling that is monitored by a photodiode (PD), or with a pair of counterpropagating beams (dashed and solid lines) for laser cooling~\cite{Kogel2024isotope}. After subsequent passage through an interaction region, which is currently $16\,$cm long but can reach multiple meters if a magnet is included to enhance parity-violating effects~\cite{Altuntas2018}, the molecular distribution is detected using laser-induced fluorescence on an EMCCD camera. Dimensions are not to scale.}
    \label{fig:setup}
\end{figure}

\textit{Experimental setup---} Our setup to demonstrate transverse laser cooling in \baff{137} is based on a standard molecular beam apparatus~\cite{Albrecht2020,Rockenhaeuser2023} similar to the one used for parity violation searches in the \baff{138} test system~\cite{Altuntas2018}, and in several ongoing electron electric dipole moment searches~\cite{Hudson2011,ACME2018,Aggarwal2018} (see Fig.~\ref{fig:setup}). In short, we create a molecular beam containing all stable isotopologues of BaF using a $3.5\,$K cryogenic buffer gas source. The molecules leaving the source have a forward velocity of around $150\,$ m/s and are collimated by apertures to transverse velocities of around $\pm2.5\,$m/s. They subsequently enter a $22\,$cm long cooling region where they interact with a pair of counterpropagating laser beams, which are retro-reflected $17$ times to form an intensity standing wave pattern. In an actual beam measurement of parity violation, this laser cooling region would be followed by the meter-long interaction region discussed above. In our proof-of-principle experiment, the molecules instead expand freely for $16\,$cm, before they are detected via laser-induced fluorescence using an EMCCD camera.

\begin{figure*}[tb]
    \centering
\includegraphics[]{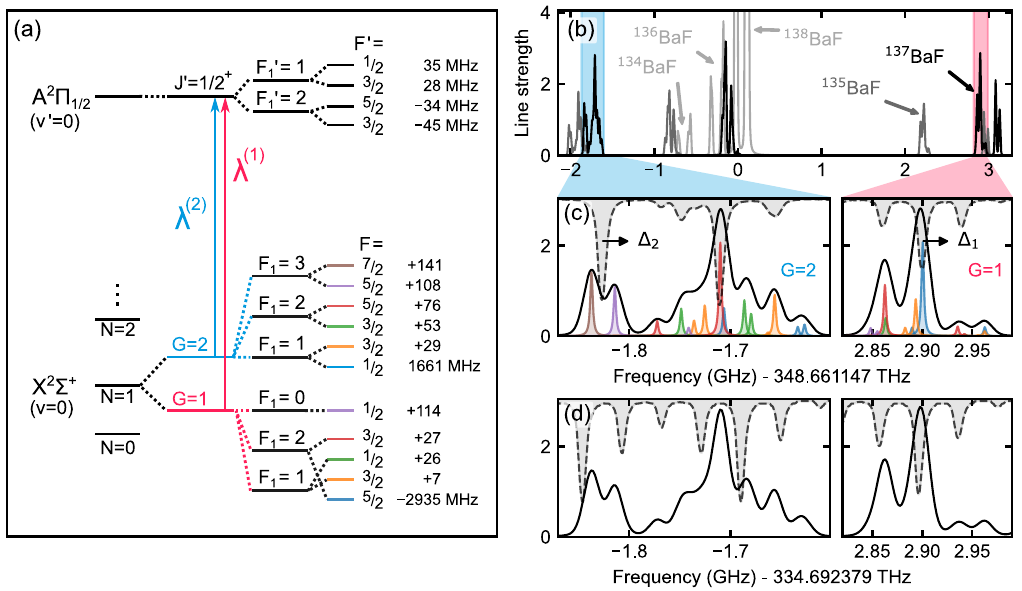}   
    \caption{(a) Level structure of \baff{137} molecules, with $N$, $J$, $G$, $F_1$ and $F$ denoting quantum numbers related to the angular momenta described in the text. (b) Simulated spectrum including the spectral lines of the other naturally occurring isotopologues of BaF. (c,d) Zoom into the shaded regions from (b), corresponding to the $G=1$ and $G=2$ hyperfine manifolds, which are addressed with laser wavelengths $\lambda^{(1)}$ and $\lambda^{(2)}$ in (a). (c) Transitions with natural linewidth (solid lines, bottom) and corresponding sidebands (dashed line, top) used to address these transitions for optical cycling and laser cooling. Transitions with the same color have the same hyperfine ground state. The solid black line is the sum of the individual molecular transitions including Doppler broadening of $7\,$ MHz. For optical cycling the sidebands are applied resonantly, as shown. For cooling the full  spectra are shifted to realize detunings $\Delta_1$ and $\Delta_2$. The frequency axis is given relative to the absolute frequency of the transition. (d) Same, but showing the conventional sinusoidal sidebands used for repumping. Nearly identical sidebands are also used for depumping during detection.}
    \label{fig:levelscheme}
\end{figure*}

\textit{Closed cycling transition---}The principles of laser cooling and trapping of various even BaF isotopologues have recently been demonstrated in a series of experiments~\cite{Rockenhaeuser2024,Kogel2024isotope,Kogel2024serrodynes,Zeng2024}. However, as discussed above, these isotopologues do not show any relevant parity violation. Instead, odd isotopologues are required, where the barium nuclear spin does not vanish. This leads to complex hyperfine spectra, for which laser cooling has so far only been proposed~\cite{Kogel2021}, but not yet been realized.

Despite the greater complexity of the level structure in odd \baff{137} compared to even \baff{138}, the overall properties that make BaF a favorable species for laser cooling remain intact. The Franck-Condon factors for vibrational transitions are highly diagonal, minimizing losses to higher vibrational states. Rotational branching can be suppressed using parity selection rules, by driving the usual transition between the lowest odd-parity ground state \gs $(N=1)$ and the lowest even-parity exited state \exs$(J'=1/2^+)$~\cite{Fitch2021,Albrecht2020}. Here, $N$ denotes the rotational quantum number in the ground state and $J'$ is the total angular momentum excluding nuclear spin in the excited state. Although this selection rule is only approximately valid in the odd isotopologues of BaF due to increased excited-state hyperfine mixing, it allows for the scattering of approximately $2000$ photons, before significant branching into higher rotational states occurs~\cite{Kogel2024fermispectroscopy}. This is sufficient for efficient transverse laser cooling~\cite{Rockenhaeuser2024}.

The relevant level structure of \baff{137} is summarized in Fig.~\ref{fig:levelscheme}a~\cite{Steimle2011,Kogel2024fermispectroscopy}. This level structure is strongly influenced by the two nuclear spins, with the fluorine atom carrying $I_\mathrm{F}=1/2$ and the odd barium atom contributing $I_{{}^{137}\mathrm{Ba}}=3/2$. In the ground state \gs, the spin $\mathbf{S}$ of the single free valence electron of the molecules strongly couples to the barium nuclear spin to form an intermediate angular momentum $\mathbf{G} = \mathbf{S} + \mathbf{I}_{{}^{137}\mathrm{Ba}}$, resulting in two well-separated hyperfine manifolds with quantum numbers $G = 1$ and $G = 2$. These manifolds are split by approximately $4.6\,$GHz. Further coupling with the rotational angular momentum $\mathbf{N}$ leads to $\mathbf{F}_1 = \mathbf{N} + \mathbf{G}$, followed by coupling with the fluorine nuclear spin to form the total angular momentum $\mathbf{F} = \mathbf{F}_1 + \mathbf{I}_\mathrm{F}$. All resulting hyperfine states, both from the barium and fluorine induced hyperfine interactions, are well resolved and must be properly addressed to avoid dark states. In the excited state \exs, the total angular momentum $\mathbf{J}$ is sequentially coupled with $\mathbf{I}_{{}^{137}\mathrm{Ba}}$ and $\mathbf{I}_\mathrm{F}$ to form $\mathbf{F}_1$ and $\mathbf{F}$, respectively. Again, both of these hyperfine splittings are resolved in BaF molecules~\cite{Rockenhaeuser2023}. In total, the ground state consists of $11$, and the excited state of $4$ hyperfine states, each containing many more magnetic sublevels. These states are connected by $35$ allowed transitions that cover a wide range of frequencies~\cite{Kogel2024fermispectroscopy}, making it difficult to address the molecules using conventional sideband schemes. 

Instead, we employ serrodynes to tailor optimized optical spectra, which can be chosen to match the molecular spectrum for optical cycling and fluorescence detection precisely or selectively address the transitions expected to contribute the most to laser cooling forces~\cite{Rockenhaeuser2024,Kogel2024serrodynes}. In both cases, overlap with any undesired nearby transitions can be effectively avoided, while the scattering rate and cooling forces are maximized, respectively.

To create the light to manipulate the molecules, we use various diode lasers matched to the respective cooling ($\nu=0\rightarrow\nu'=0$), repumping ($\nu=1\rightarrow\nu'=0$) and depumping ($\nu=0\rightarrow\nu'=1$) transitions for BaF, where $\nu$ and $\nu'$ denote the vibrational quantum numbers in the ground and excited states, respectively. All lasers used are stabilized with MHz precision by a transfer lock~\cite{Pultinevicius2023}, modulated by suitable free space and fiber-coupled electro-optical modulators (EOMs) to be detailed further below, and finally amplified using tapered amplifiers~\cite{Kogel2024serrodynes}. Since the significant splitting between the hyperfine manifolds $G=1$ and $G=2$ increases the total number of laser systems required to address all molecular states, only the first vibrational repumping laser is implemented in the present work. More repumping lasers could easily be added along the same lines, reducing losses through higher-order vibrational branching. 

\begin{figure}[tb]
    \centering
    \includegraphics[]{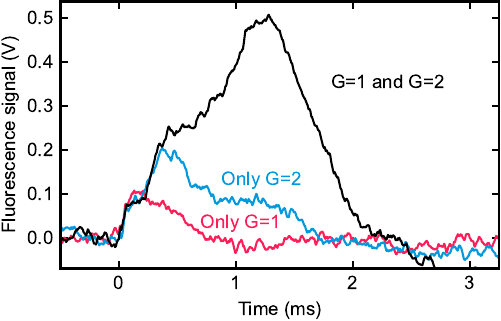}
    \caption{Optical cycling of \baff{137}. Laser-induced fluorescence of the molecules passing a single transversal laser beam. The fluorescence increases strongly from the quasi-closed optical cycling transition realized when both the $G=1$ and $G=2$ hyperfine manifolds are addressed, compared to when only one of the two is addressed and molecules are quickly lost into dark states. The observed individual signal strengths for $G=1$ and $G=2$ are consistent with the respective number of ground state levels and the branching ratios into the manifolds from the excited state.}
    \label{fig:opticalcycling}
\end{figure}

\textit{Optical Cycling---}In a first step, we realize optical cycling of the molecules in the laser cooling region by detecting the change of fluorescence signal when closing the optical cycle. Such optical cycling is a prerequisite for laser cooling and is also essential for efficient state preparation and detection~\cite{Altuntas2018,Lasner2018}. 

To probe its properties, we use a single laser beam with a diameter of $3.1\,$mm full width at half maximum in the cooling region, blocking the second beam to suppress the resulting standing wave pattern responsible for laser cooling. We employ a comparably low power of $8.4\,$mW per hyperfine manifold and apply a magnetic field of $1\,$G, rotated by $45^\circ$ relative to the laser polarization axis, to remix dark magnetic sublevels~\cite{Fitch2021}. The sideband configuration used is depicted in Fig.~\ref{fig:levelscheme}c. We match the complex molecular spectrum of the $G=2$ manifold using a serrodyne waveform applied via a fiber-coupled EOM, while for the simpler $G=1$ manifold a second laser with a sinusoidally driven $39.3\,$MHz free-space EOM is employed. This frequency is chosen because it most closely matches the relevant hyperfine substructure induced by the fluorine nuclear spin. The use of separate lasers ensures that ample and freely tunable power is available for each hyperfine manifold.

The resulting fluorescence signals are recorded using an amplified photodiode and are shown in Fig.~\ref{fig:opticalcycling}. When only one of the $G=1$ and $G=2$ hyperfine manifolds is addressed individually, the number of photons emitted by a molecule is limited to $1-3$ by strong branching into the respective other manifold. However, when the laser systems addressing both manifolds are used in parallel, this leakage is effectively suppressed and a quasi-closed optical cycle is formed, as highlighted by a strong increase in fluorescence signal. 

\textit{Transverse laser cooling---} In the next step, we realize transverse laser cooling of the molecular beam. For this, we apply both cooling and repumping light throughout the entire cooling region, with a power of $120\,$mW each, addressing a total of $112$  molecular magnetic sublevels. A $1\,$G magnetic field destabilizes those dark sublevels that do not directly couple to the laser light. 

 Informed by our previous work on optimizing the spectra for cooling of the simpler \baff{138}~\cite{Rockenhaeuser2024,Kogel2024serrodynes}, for the $G=2$ manifold we use serrodyne spectra matching the molecular spectra, similar to those used before for optical cycling. We specifically enhance the strength of the sidebands that address the strongest transitions with the highest multiplicities in the ground state. These are expected to contribute the most to the laser cooling force, while other transitions only need to be driven weakly to avoid dark states. For $G=1$ we again employ a separate laser combined with a sinusoidally driven free-space EOM, which is sufficient to address dark states in this manifold. For cooling, we shift both spectra relative to the respective molecular spectra by individual detunings $\Delta_1$ and $\Delta_2$. Although this configuration does not realize the strongest forces possible, it represents the most direct optimization strategy possible for complex molecules. It does not require precise spectroscopic knowledge of the complex spectra and thus provides a robust way to realize cooling of many molecular species~\cite{Kogel2024serrodynes}.

 \begin{figure}[tb]
    \centering
    \includegraphics[]{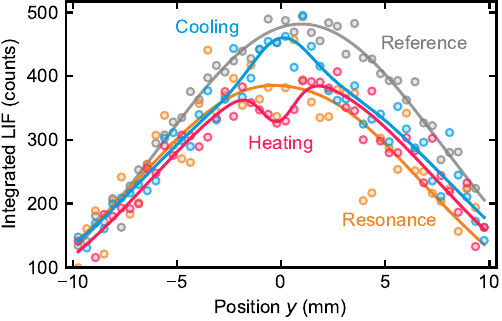}
    \caption{Laser cooling of \baff{137}, showing the typical signatures of magnetically-assisted Sisyphus cooling. For suitable sideband detunings, molecules either accumulate in the coldest part of the molecular beam (blue) or are heated away from it (red). No change is observed for resonance light (orange). For all signals, the presence of cooling light leads to decay into unaddressed higher vibrational levels, resulting in an overall decrease compared to the reference where no lasers are applied at all (gray). As in previous work~\cite{Kogel2024isotope}, this decay is purely technical and could readily be removed with additional repumping lasers. Datapoints are averaged over $400$ realizations and solid lines are double Gaussian fits~\cite{Rockenhaeuser2024}.}
    \label{fig:lasercooling}
\end{figure}

 In addition, we also use two lasers for repumping to address both manifolds. These are combined and both modulated using
a single sinusoidally driven $39.3\,$MHz free-space EOM that strongly modulates sidebands up to the third order (see Fig.~\ref{fig:levelscheme}d). For detection, we employ Raman optical cycling~\cite{Shaw2021,Rockenhaeuser2023}, where a depumping laser is combined with an additional repumping laser to facilitate a high signal-to-noise ratio and isotopologue selectivity~\cite{Kogel2024isotope}. For depumping, we use a single laser with a sinusoidally driven free-space EOM and a second $2.314\,$GHz free-space EOM, demonstrating an alternative for addressing both hyperfine manifolds simultaneously. This approach results in a spectrum nearly identical to that shown for repumping in Fig.~\ref{fig:levelscheme}d.

To characterize the cooling forces, we blue-detune $\Delta_1$ by $+20\,$MHz, and use $\Delta_2=-12, 0,$ and $+15\,$MHz to tune between red-detuned heating, resonant scattering, and blue-detuned cooling. The results are shown in Fig.~\ref{fig:lasercooling}. We observe the characteristic signatures of magnetically-assisted Sisyphus laser cooling---accumulation of molecules in the coldest region of the beam for blue detuning, whereas molecules are repelled from this region for red detuning. We find that the enhancement in the coldest part of the molecular beam is comparable to the results reported for other heavy molecules~\cite{Lim2018,Kozyryev2017}, highlighting forces strong enough to  collimate the beam to efficiently pass through a suitable interaction region. From a double Gaussian fit, we extract the peak height ratio between the cooled and uncooled molecules~\cite{Rockenhaeuser2024}, obtaining a value of $0.15$. From this we conservatively estimate that the cooling forces realized here are approximately $2$ times lower than the forces realized in \baff{138} when using the same basic sideband optimization strategy. As discussed above and in Ref.~\cite{Kogel2024serrodynes}, we expect that a more comprehensive optimization, e.g., by allowing freely tunable sidebands in both hyperfine manifolds or by changing the total number of frequency components, will further enhance these forces and make them comparable to those observed in \baff{138}.

\textit{Conclusion---} We have demonstrated optical cycling and laser cooling of a \baff{137} molecular beam. 

The optical cycling techniques presented have the potential for immediate improvements in searches for nuclear-spin-dependent parity violation using these molecules, through more efficient target science state preparation and detection. As the initial population in \baff{137}---compared to \baff{138}---is spread over many possible hyperfine states, optical pumping realized by removing specific sidebands (see Fig.~\ref{fig:levelscheme}c) could significantly increase the signal strength in existing setups~\cite{Altuntas2018}.

In terms of laser cooling, even for the comparatively straightforward transversal cooling demonstrated in this work, the realization of the closed optical cycle required addressing a number of molecular levels that exceed the number present in the most complex polyatomics cooled to date~\cite{Mitra2020,Vilas2021}. Our results thus open up a new class of molecules with complex hyperfine structure, which have notable applications in many types of precision measurements~\cite{Kozlov1995,DeMille2008,Kudashov2014,Norrgard2019,Hao2020, Hutzler2020,Jadbabaie2023}. Including transverse focusing~\cite{Touwen2024} or a second transverse cooling direction~\cite{Alauze2021} is straightforward and would increase the flux of a \baff{137} molecular beam by many orders of magnitude~\cite{Kogel2021}. This promises significant improvements over the best atomic measurements of parity violation, with potentially important implications for the extraction and interpretation of SM and beyond SM physics~\cite{Safronova2018, Johnson2003,Dzuba2017,Flambaum1997,Haxton2002}. 

The techniques presented will also be transferable to a magneto-optical trap (MOT), as recently realized for the simpler \baff{138}~\cite{Zeng2024}. In this realization, all relevant loss channels currently expected to be present in \baff{137} have already been addressed. In addition to more vibrational repumping lasers, similar to the situation in yttrium monoxide~\cite{Collopy2018}, such a MOT will require the remixing of additional rotational levels due to branching losses via an intermediate electronic $A'\Delta$ state, resulting in a prospective quasi-closed optical cycle including around $750$ magnetic sublevels. Although this number is large and far beyond the complexity in any previously trapped molecule, we see no fundamental limitations preventing the realization of such a MOT.

As the demonstrated laser cooling techniques are highly specific to a particular isotopologue, the realization of a MOT could facilitate systematic parity-violation measurements across several rare isotopologues~\cite{Antypas2019}, e.g. \baff{135} and radioactive \baff{133}~\cite{Hoehle1976}, or chains of isotopologues in other similar molecules like YbF, BaOH, YbOH or RaF~\cite{Denis2020,Zeng2023,Udrescu2021}. This would allow differential measurements to be made that could disentangle the competing parameters that encode the properties of the weak interaction without the need for challenging molecular structure calculations, which are typically subject to significant uncertainties. Furthermore, such systematic measurements could also serve as benchmarks for nuclear calculations of anapole moments, which are notoriously challenging in heavy nuclei~\cite{Haxton2013}. Taken together, this could open a new window into the weak interaction and its effects in nuclear and particle physics.

\section*{Acknowledgments}
We are indebted to Tilman Pfau for generous support and thank Anastasia Borschevsky, Yuly Chamorro, Dave DeMille, Mangesh Bhattarai and Richard Mawhorter for discussions and sharing preliminary data on the structure of odd BaF isotopologues. This project has received funding from the European Research Council (ERC) under the European Union’s Horizon 2020 research and innovation program (Grant agreement No. 949431), Vector Stiftung, Carl Zeiss Stiftung, the RiSC initiative of the Ministry of Science, Research and Arts Baden-W\"urttemberg, and was funded in whole or in part by the Austrian Science Fund (FWF) 10.55776/PAT8306623.

\bibliography{biblio}

\begin{thebibliography}{63}%
\makeatletter
\providecommand \@ifxundefined [1]{%
 \@ifx{#1\undefined}
}%
\providecommand \@ifnum [1]{%
 \ifnum #1\expandafter \@firstoftwo
 \else \expandafter \@secondoftwo
 \fi
}%
\providecommand \@ifx [1]{%
 \ifx #1\expandafter \@firstoftwo
 \else \expandafter \@secondoftwo
 \fi
}%
\providecommand \natexlab [1]{#1}%
\providecommand \enquote  [1]{``#1''}%
\providecommand \bibnamefont  [1]{#1}%
\providecommand \bibfnamefont [1]{#1}%
\providecommand \citenamefont [1]{#1}%
\providecommand \href@noop [0]{\@secondoftwo}%
\providecommand \href [0]{\begingroup \@sanitize@url \@href}%
\providecommand \@href[1]{\@@startlink{#1}\@@href}%
\providecommand \@@href[1]{\endgroup#1\@@endlink}%
\providecommand \@sanitize@url [0]{\catcode `\\12\catcode `\$12\catcode `\&12\catcode `\#12\catcode `\^12\catcode `\_12\catcode `\%12\relax}%
\providecommand \@@startlink[1]{}%
\providecommand \@@endlink[0]{}%
\providecommand \url  [0]{\begingroup\@sanitize@url \@url }%
\providecommand \@url [1]{\endgroup\@href {#1}{\urlprefix }}%
\providecommand \urlprefix  [0]{URL }%
\providecommand \Eprint [0]{\href }%
\providecommand \doibase [0]{https://doi.org/}%
\providecommand \selectlanguage [0]{\@gobble}%
\providecommand \bibinfo  [0]{\@secondoftwo}%
\providecommand \bibfield  [0]{\@secondoftwo}%
\providecommand \translation [1]{[#1]}%
\providecommand \BibitemOpen [0]{}%
\providecommand \bibitemStop [0]{}%
\providecommand \bibitemNoStop [0]{.\EOS\space}%
\providecommand \EOS [0]{\spacefactor3000\relax}%
\providecommand \BibitemShut  [1]{\csname bibitem#1\endcsname}%
\let\auto@bib@innerbib\@empty
\bibitem [{\citenamefont {Hudson}\ \emph {et~al.}(2011)\citenamefont {Hudson}, \citenamefont {Kara}, \citenamefont {Smallman}, \citenamefont {Sauer}, \citenamefont {Tarbutt},\ and\ \citenamefont {Hinds}}]{Hudson2011}%
  \BibitemOpen
  \bibfield  {author} {\bibinfo {author} {\bibfnamefont {J.~J.}\ \bibnamefont {Hudson}}, \bibinfo {author} {\bibfnamefont {D.~M.}\ \bibnamefont {Kara}}, \bibinfo {author} {\bibfnamefont {I.~J.}\ \bibnamefont {Smallman}}, \bibinfo {author} {\bibfnamefont {B.~E.}\ \bibnamefont {Sauer}}, \bibinfo {author} {\bibfnamefont {M.~R.}\ \bibnamefont {Tarbutt}},\ and\ \bibinfo {author} {\bibfnamefont {E.~A.}\ \bibnamefont {Hinds}},\ }\bibfield  {title} {\bibinfo {title} {Improved measurement of the shape of the electron},\ }\href {https://doi.org/10.1038/nature10104} {\bibfield  {journal} {\bibinfo  {journal} {Nature}\ }\textbf {\bibinfo {volume} {473}},\ \bibinfo {pages} {493} (\bibinfo {year} {2011})}\BibitemShut {NoStop}%
\bibitem [{\citenamefont {Andreev}\ \emph {et~al.}(2018)\citenamefont {Andreev}, \citenamefont {Ang}, \citenamefont {DeMille}, \citenamefont {Doyle}, \citenamefont {Gabrielse}, \citenamefont {Haefner}, \citenamefont {Hutzler}, \citenamefont {Lasner}, \citenamefont {Meisenhelder}, \citenamefont {O'Leary}, \citenamefont {Panda}, \citenamefont {West}, \citenamefont {West}, \citenamefont {Wu},\ and\ \citenamefont {Collaboration}}]{ACME2018}%
  \BibitemOpen
  \bibfield  {author} {\bibinfo {author} {\bibfnamefont {V.}~\bibnamefont {Andreev}}, \bibinfo {author} {\bibfnamefont {D.~G.}\ \bibnamefont {Ang}}, \bibinfo {author} {\bibfnamefont {D.}~\bibnamefont {DeMille}}, \bibinfo {author} {\bibfnamefont {J.~M.}\ \bibnamefont {Doyle}}, \bibinfo {author} {\bibfnamefont {G.}~\bibnamefont {Gabrielse}}, \bibinfo {author} {\bibfnamefont {J.}~\bibnamefont {Haefner}}, \bibinfo {author} {\bibfnamefont {N.~R.}\ \bibnamefont {Hutzler}}, \bibinfo {author} {\bibfnamefont {Z.}~\bibnamefont {Lasner}}, \bibinfo {author} {\bibfnamefont {C.}~\bibnamefont {Meisenhelder}}, \bibinfo {author} {\bibfnamefont {B.~R.}\ \bibnamefont {O'Leary}}, \bibinfo {author} {\bibfnamefont {C.~D.}\ \bibnamefont {Panda}}, \bibinfo {author} {\bibfnamefont {A.~D.}\ \bibnamefont {West}}, \bibinfo {author} {\bibfnamefont {E.~P.}\ \bibnamefont {West}}, \bibinfo {author} {\bibfnamefont {X.}~\bibnamefont {Wu}},\ and\ \bibinfo {author} {\bibfnamefont {A.}~\bibnamefont {Collaboration}},\ }\bibfield  {title}
  {\bibinfo {title} {Improved limit on the electric dipole moment of the electron},\ }\href {https://doi.org/10.1038/s41586-018-0599-8} {\bibfield  {journal} {\bibinfo  {journal} {Nature}\ }\textbf {\bibinfo {volume} {562}},\ \bibinfo {pages} {355} (\bibinfo {year} {2018})}\BibitemShut {NoStop}%
\bibitem [{\citenamefont {Roussy}\ \emph {et~al.}(2023)\citenamefont {Roussy}, \citenamefont {Caldwell}, \citenamefont {Wright}, \citenamefont {Cairncross}, \citenamefont {Shagam}, \citenamefont {Ng}, \citenamefont {Schlossberger}, \citenamefont {Park}, \citenamefont {Wang}, \citenamefont {Ye},\ and\ \citenamefont {Cornell}}]{Roussy2023}%
  \BibitemOpen
  \bibfield  {author} {\bibinfo {author} {\bibfnamefont {T.~S.}\ \bibnamefont {Roussy}}, \bibinfo {author} {\bibfnamefont {L.}~\bibnamefont {Caldwell}}, \bibinfo {author} {\bibfnamefont {T.}~\bibnamefont {Wright}}, \bibinfo {author} {\bibfnamefont {W.~B.}\ \bibnamefont {Cairncross}}, \bibinfo {author} {\bibfnamefont {Y.}~\bibnamefont {Shagam}}, \bibinfo {author} {\bibfnamefont {K.~B.}\ \bibnamefont {Ng}}, \bibinfo {author} {\bibfnamefont {N.}~\bibnamefont {Schlossberger}}, \bibinfo {author} {\bibfnamefont {S.~Y.}\ \bibnamefont {Park}}, \bibinfo {author} {\bibfnamefont {A.}~\bibnamefont {Wang}}, \bibinfo {author} {\bibfnamefont {J.}~\bibnamefont {Ye}},\ and\ \bibinfo {author} {\bibfnamefont {E.~A.}\ \bibnamefont {Cornell}},\ }\bibfield  {title} {\bibinfo {title} {An improved bound on the electron’s electric dipole moment},\ }\href {https://doi.org/10.1126/science.adg4084} {\bibfield  {journal} {\bibinfo  {journal} {Science}\ }\textbf {\bibinfo {volume} {381}},\ \bibinfo {pages} {46} (\bibinfo {year}
  {2023})}\BibitemShut {NoStop}%
\bibitem [{\citenamefont {Shelkovnikov}\ \emph {et~al.}(2008)\citenamefont {Shelkovnikov}, \citenamefont {Butcher}, \citenamefont {Chardonnet},\ and\ \citenamefont {Amy-Klein}}]{Shelkovnikov2008}%
  \BibitemOpen
  \bibfield  {author} {\bibinfo {author} {\bibfnamefont {A.}~\bibnamefont {Shelkovnikov}}, \bibinfo {author} {\bibfnamefont {R.~J.}\ \bibnamefont {Butcher}}, \bibinfo {author} {\bibfnamefont {C.}~\bibnamefont {Chardonnet}},\ and\ \bibinfo {author} {\bibfnamefont {A.}~\bibnamefont {Amy-Klein}},\ }\bibfield  {title} {\bibinfo {title} {Stability of the proton-to-electron mass ratio},\ }\href {https://doi.org/10.1103/PhysRevLett.100.150801} {\bibfield  {journal} {\bibinfo  {journal} {Phys. Rev. Lett.}\ }\textbf {\bibinfo {volume} {100}},\ \bibinfo {pages} {150801} (\bibinfo {year} {2008})}\BibitemShut {NoStop}%
\bibitem [{\citenamefont {Chin}\ \emph {et~al.}(2009)\citenamefont {Chin}, \citenamefont {Flambaum},\ and\ \citenamefont {Kozlov}}]{Chin2009}%
  \BibitemOpen
  \bibfield  {author} {\bibinfo {author} {\bibfnamefont {C.}~\bibnamefont {Chin}}, \bibinfo {author} {\bibfnamefont {V.~V.}\ \bibnamefont {Flambaum}},\ and\ \bibinfo {author} {\bibfnamefont {M.~G.}\ \bibnamefont {Kozlov}},\ }\bibfield  {title} {\bibinfo {title} {Ultracold molecules: new probes on the variation of fundamental constants},\ }\href {https://doi.org/10.1088/1367-2630/11/5/055048} {\bibfield  {journal} {\bibinfo  {journal} {New Journal of Physics}\ }\textbf {\bibinfo {volume} {11}},\ \bibinfo {pages} {055048} (\bibinfo {year} {2009})}\BibitemShut {NoStop}%
\bibitem [{\citenamefont {Truppe}\ \emph {et~al.}(2017)\citenamefont {Truppe}, \citenamefont {Williams}, \citenamefont {Hambach}, \citenamefont {Caldwell}, \citenamefont {Fitch}, \citenamefont {Hinds}, \citenamefont {Sauer},\ and\ \citenamefont {Tarbutt}}]{Truppe2017}%
  \BibitemOpen
  \bibfield  {author} {\bibinfo {author} {\bibfnamefont {S.}~\bibnamefont {Truppe}}, \bibinfo {author} {\bibfnamefont {H.~J.}\ \bibnamefont {Williams}}, \bibinfo {author} {\bibfnamefont {M.}~\bibnamefont {Hambach}}, \bibinfo {author} {\bibfnamefont {L.}~\bibnamefont {Caldwell}}, \bibinfo {author} {\bibfnamefont {N.~J.}\ \bibnamefont {Fitch}}, \bibinfo {author} {\bibfnamefont {E.~A.}\ \bibnamefont {Hinds}}, \bibinfo {author} {\bibfnamefont {B.~E.}\ \bibnamefont {Sauer}},\ and\ \bibinfo {author} {\bibfnamefont {M.~R.}\ \bibnamefont {Tarbutt}},\ }\bibfield  {title} {\bibinfo {title} {{Molecules cooled below the Doppler limit}},\ }\href {https://doi.org/10.1038/nphys4241} {\bibfield  {journal} {\bibinfo  {journal} {Nature Physics}\ }\textbf {\bibinfo {volume} {13}},\ \bibinfo {pages} {1173} (\bibinfo {year} {2017})}\BibitemShut {NoStop}%
\bibitem [{\citenamefont {Kobayashi}\ \emph {et~al.}(2019)\citenamefont {Kobayashi}, \citenamefont {Ogino},\ and\ \citenamefont {Inouye}}]{Kobayashi2019}%
  \BibitemOpen
  \bibfield  {author} {\bibinfo {author} {\bibfnamefont {J.}~\bibnamefont {Kobayashi}}, \bibinfo {author} {\bibfnamefont {A.}~\bibnamefont {Ogino}},\ and\ \bibinfo {author} {\bibfnamefont {S.}~\bibnamefont {Inouye}},\ }\bibfield  {title} {\bibinfo {title} {Measurement of the variation of electron-to-proton mass ratio using ultracold molecules produced from laser-cooled atoms},\ }\href {https://doi.org/10.1038/s41467-019-11761-1} {\bibfield  {journal} {\bibinfo  {journal} {Nature Communications}\ }\textbf {\bibinfo {volume} {10}},\ \bibinfo {pages} {3771} (\bibinfo {year} {2019})}\BibitemShut {NoStop}%
\bibitem [{\citenamefont {Alighanbari}\ \emph {et~al.}(2020)\citenamefont {Alighanbari}, \citenamefont {Giri}, \citenamefont {Constantin}, \citenamefont {Korobov},\ and\ \citenamefont {Schiller}}]{Alighanbari2020}%
  \BibitemOpen
  \bibfield  {author} {\bibinfo {author} {\bibfnamefont {S.}~\bibnamefont {Alighanbari}}, \bibinfo {author} {\bibfnamefont {G.~S.}\ \bibnamefont {Giri}}, \bibinfo {author} {\bibfnamefont {F.~L.}\ \bibnamefont {Constantin}}, \bibinfo {author} {\bibfnamefont {V.~I.}\ \bibnamefont {Korobov}},\ and\ \bibinfo {author} {\bibfnamefont {S.}~\bibnamefont {Schiller}},\ }\bibfield  {title} {\bibinfo {title} {Precise test of quantum electrodynamics and determination of fundamental constants with {HD+} ions},\ }\href {https://doi.org/10.1038/s41586-020-2261-5} {\bibfield  {journal} {\bibinfo  {journal} {Nature}\ }\textbf {\bibinfo {volume} {581}},\ \bibinfo {pages} {152} (\bibinfo {year} {2020})}\BibitemShut {NoStop}%
\bibitem [{\citenamefont {Germann}\ \emph {et~al.}(2021)\citenamefont {Germann}, \citenamefont {Patra}, \citenamefont {Karr}, \citenamefont {Hilico}, \citenamefont {Korobov}, \citenamefont {Salumbides}, \citenamefont {Eikema}, \citenamefont {Ubachs},\ and\ \citenamefont {Koelemeij}}]{Germann2021}%
  \BibitemOpen
  \bibfield  {author} {\bibinfo {author} {\bibfnamefont {M.}~\bibnamefont {Germann}}, \bibinfo {author} {\bibfnamefont {S.}~\bibnamefont {Patra}}, \bibinfo {author} {\bibfnamefont {J.-P.}\ \bibnamefont {Karr}}, \bibinfo {author} {\bibfnamefont {L.}~\bibnamefont {Hilico}}, \bibinfo {author} {\bibfnamefont {V.~I.}\ \bibnamefont {Korobov}}, \bibinfo {author} {\bibfnamefont {E.~J.}\ \bibnamefont {Salumbides}}, \bibinfo {author} {\bibfnamefont {K.~S.~E.}\ \bibnamefont {Eikema}}, \bibinfo {author} {\bibfnamefont {W.}~\bibnamefont {Ubachs}},\ and\ \bibinfo {author} {\bibfnamefont {J.~C.~J.}\ \bibnamefont {Koelemeij}},\ }\bibfield  {title} {\bibinfo {title} {Three-body qed test and fifth-force constraint from vibrations and rotations of {${\mathrm{HD}}^{+}$}},\ }\href {https://doi.org/10.1103/PhysRevResearch.3.L022028} {\bibfield  {journal} {\bibinfo  {journal} {Phys. Rev. Res.}\ }\textbf {\bibinfo {volume} {3}},\ \bibinfo {pages} {L022028} (\bibinfo {year} {2021})}\BibitemShut {NoStop}%
\bibitem [{\citenamefont {Doran}\ \emph {et~al.}(2024)\citenamefont {Doran}, \citenamefont {H\"olsch}, \citenamefont {Beyer},\ and\ \citenamefont {Merkt}}]{Doran2024}%
  \BibitemOpen
  \bibfield  {author} {\bibinfo {author} {\bibfnamefont {I.}~\bibnamefont {Doran}}, \bibinfo {author} {\bibfnamefont {N.}~\bibnamefont {H\"olsch}}, \bibinfo {author} {\bibfnamefont {M.}~\bibnamefont {Beyer}},\ and\ \bibinfo {author} {\bibfnamefont {F.}~\bibnamefont {Merkt}},\ }\bibfield  {title} {\bibinfo {title} {Zero-quantum-defect method and the fundamental vibrational interval of ${{\mathrm{H}}}_{2}^{+}$},\ }\href {https://doi.org/10.1103/PhysRevLett.132.073001} {\bibfield  {journal} {\bibinfo  {journal} {Phys. Rev. Lett.}\ }\textbf {\bibinfo {volume} {132}},\ \bibinfo {pages} {073001} (\bibinfo {year} {2024})}\BibitemShut {NoStop}%
\bibitem [{\citenamefont {DeMille}\ \emph {et~al.}(2017)\citenamefont {DeMille}, \citenamefont {Doyle},\ and\ \citenamefont {Sushkov}}]{DeMille2017}%
  \BibitemOpen
  \bibfield  {author} {\bibinfo {author} {\bibfnamefont {D.}~\bibnamefont {DeMille}}, \bibinfo {author} {\bibfnamefont {J.~M.}\ \bibnamefont {Doyle}},\ and\ \bibinfo {author} {\bibfnamefont {A.~O.}\ \bibnamefont {Sushkov}},\ }\bibfield  {title} {\bibinfo {title} {{Probing the frontiers of particle physics with tabletop-scale experiments}},\ }\href {https://doi.org/10.1126/science.aal3003} {\bibfield  {journal} {\bibinfo  {journal} {Science}\ }\textbf {\bibinfo {volume} {357}},\ \bibinfo {pages} {990} (\bibinfo {year} {2017})}\BibitemShut {NoStop}%
\bibitem [{\citenamefont {Safronova}\ \emph {et~al.}(2018)\citenamefont {Safronova}, \citenamefont {Budker}, \citenamefont {DeMille}, \citenamefont {Kimball}, \citenamefont {Derevianko},\ and\ \citenamefont {Clark}}]{Safronova2018}%
  \BibitemOpen
  \bibfield  {author} {\bibinfo {author} {\bibfnamefont {M.~S.}\ \bibnamefont {Safronova}}, \bibinfo {author} {\bibfnamefont {D.}~\bibnamefont {Budker}}, \bibinfo {author} {\bibfnamefont {D.}~\bibnamefont {DeMille}}, \bibinfo {author} {\bibfnamefont {D.~F.~J.}\ \bibnamefont {Kimball}}, \bibinfo {author} {\bibfnamefont {A.}~\bibnamefont {Derevianko}},\ and\ \bibinfo {author} {\bibfnamefont {C.~W.}\ \bibnamefont {Clark}},\ }\bibfield  {title} {\bibinfo {title} {Search for new physics with atoms and molecules},\ }\href {https://doi.org/10.1103/RevModPhys.90.025008} {\bibfield  {journal} {\bibinfo  {journal} {Rev. Mod. Phys.}\ }\textbf {\bibinfo {volume} {90}},\ \bibinfo {pages} {025008} (\bibinfo {year} {2018})}\BibitemShut {NoStop}%
\bibitem [{\citenamefont {DeMille}\ \emph {et~al.}(2024)\citenamefont {DeMille}, \citenamefont {Hutzler}, \citenamefont {Rey},\ and\ \citenamefont {Zelevinsky}}]{Demille2024}%
  \BibitemOpen
  \bibfield  {author} {\bibinfo {author} {\bibfnamefont {D.}~\bibnamefont {DeMille}}, \bibinfo {author} {\bibfnamefont {N.~R.}\ \bibnamefont {Hutzler}}, \bibinfo {author} {\bibfnamefont {A.~M.}\ \bibnamefont {Rey}},\ and\ \bibinfo {author} {\bibfnamefont {T.}~\bibnamefont {Zelevinsky}},\ }\bibfield  {title} {\bibinfo {title} {Quantum sensing and metrology for fundamental physics with molecules},\ }\href {https://doi.org/10.1038/s41567-024-02499-9} {\bibfield  {journal} {\bibinfo  {journal} {Nature Physics}\ }\textbf {\bibinfo {volume} {20}},\ \bibinfo {pages} {741} (\bibinfo {year} {2024})}\BibitemShut {NoStop}%
\bibitem [{\citenamefont {Lim}\ \emph {et~al.}(2018)\citenamefont {Lim}, \citenamefont {Almond}, \citenamefont {Trigatzis}, \citenamefont {Devlin}, \citenamefont {Fitch}, \citenamefont {Sauer}, \citenamefont {Tarbutt},\ and\ \citenamefont {Hinds}}]{Lim2018}%
  \BibitemOpen
  \bibfield  {author} {\bibinfo {author} {\bibfnamefont {J.}~\bibnamefont {Lim}}, \bibinfo {author} {\bibfnamefont {J.~R.}\ \bibnamefont {Almond}}, \bibinfo {author} {\bibfnamefont {M.~A.}\ \bibnamefont {Trigatzis}}, \bibinfo {author} {\bibfnamefont {J.~A.}\ \bibnamefont {Devlin}}, \bibinfo {author} {\bibfnamefont {N.~J.}\ \bibnamefont {Fitch}}, \bibinfo {author} {\bibfnamefont {B.~E.}\ \bibnamefont {Sauer}}, \bibinfo {author} {\bibfnamefont {M.~R.}\ \bibnamefont {Tarbutt}},\ and\ \bibinfo {author} {\bibfnamefont {E.~A.}\ \bibnamefont {Hinds}},\ }\bibfield  {title} {\bibinfo {title} {Laser cooled {YbF} molecules for measuring the electron's electric dipole moment},\ }\href {https://doi.org/10.1103/PhysRevLett.120.123201} {\bibfield  {journal} {\bibinfo  {journal} {Phys. Rev. Lett.}\ }\textbf {\bibinfo {volume} {120}},\ \bibinfo {pages} {123201} (\bibinfo {year} {2018})}\BibitemShut {NoStop}%
\bibitem [{\citenamefont {Alauze}\ \emph {et~al.}(2021)\citenamefont {Alauze}, \citenamefont {Lim}, \citenamefont {Trigatzis}, \citenamefont {Swarbrick}, \citenamefont {Collings}, \citenamefont {Fitch}, \citenamefont {Sauer},\ and\ \citenamefont {Tarbutt}}]{Alauze2021}%
  \BibitemOpen
  \bibfield  {author} {\bibinfo {author} {\bibfnamefont {X.}~\bibnamefont {Alauze}}, \bibinfo {author} {\bibfnamefont {J.}~\bibnamefont {Lim}}, \bibinfo {author} {\bibfnamefont {M.~A.}\ \bibnamefont {Trigatzis}}, \bibinfo {author} {\bibfnamefont {S.}~\bibnamefont {Swarbrick}}, \bibinfo {author} {\bibfnamefont {F.~J.}\ \bibnamefont {Collings}}, \bibinfo {author} {\bibfnamefont {N.~J.}\ \bibnamefont {Fitch}}, \bibinfo {author} {\bibfnamefont {B.~E.}\ \bibnamefont {Sauer}},\ and\ \bibinfo {author} {\bibfnamefont {M.~R.}\ \bibnamefont {Tarbutt}},\ }\bibfield  {title} {\bibinfo {title} {{An ultracold molecular beam for testing fundamental physics}},\ }\href {https://doi.org/10.1088/2058-9565/ac107e} {\bibfield  {journal} {\bibinfo  {journal} {Quantum Science and Technology}\ }\textbf {\bibinfo {volume} {6}},\ \bibinfo {pages} {44005} (\bibinfo {year} {2021})}\BibitemShut {NoStop}%
\bibitem [{\citenamefont {McNally}\ \emph {et~al.}(2020)\citenamefont {McNally}, \citenamefont {Kozyryev}, \citenamefont {Vazquez-Carson}, \citenamefont {Wenz}, \citenamefont {Wang},\ and\ \citenamefont {Zelevinsky}}]{McNally2020}%
  \BibitemOpen
  \bibfield  {author} {\bibinfo {author} {\bibfnamefont {R.~L.}\ \bibnamefont {McNally}}, \bibinfo {author} {\bibfnamefont {I.}~\bibnamefont {Kozyryev}}, \bibinfo {author} {\bibfnamefont {S.}~\bibnamefont {Vazquez-Carson}}, \bibinfo {author} {\bibfnamefont {K.}~\bibnamefont {Wenz}}, \bibinfo {author} {\bibfnamefont {T.}~\bibnamefont {Wang}},\ and\ \bibinfo {author} {\bibfnamefont {T.}~\bibnamefont {Zelevinsky}},\ }\bibfield  {title} {\bibinfo {title} {Optical cycling, radiative deflection and laser cooling of barium monohydride (${}^{138}${BaH})},\ }\href {https://doi.org/10.1088/1367-2630/aba3e9} {\bibfield  {journal} {\bibinfo  {journal} {New Journal of Physics}\ }\textbf {\bibinfo {volume} {22}},\ \bibinfo {pages} {083047} (\bibinfo {year} {2020})}\BibitemShut {NoStop}%
\bibitem [{\citenamefont {Rockenh\"auser}\ \emph {et~al.}(2024)\citenamefont {Rockenh\"auser}, \citenamefont {Kogel}, \citenamefont {Garg}, \citenamefont {Morales-Ram\'{\i}rez},\ and\ \citenamefont {Langen}}]{Rockenhaeuser2024}%
  \BibitemOpen
  \bibfield  {author} {\bibinfo {author} {\bibfnamefont {M.}~\bibnamefont {Rockenh\"auser}}, \bibinfo {author} {\bibfnamefont {F.}~\bibnamefont {Kogel}}, \bibinfo {author} {\bibfnamefont {T.}~\bibnamefont {Garg}}, \bibinfo {author} {\bibfnamefont {S.~A.}\ \bibnamefont {Morales-Ram\'{\i}rez}},\ and\ \bibinfo {author} {\bibfnamefont {T.}~\bibnamefont {Langen}},\ }\bibfield  {title} {\bibinfo {title} {Laser cooling of barium monofluoride molecules using synthesized optical spectra},\ }\href {https://doi.org/10.1103/PhysRevResearch.6.043161} {\bibfield  {journal} {\bibinfo  {journal} {Phys. Rev. Res.}\ }\textbf {\bibinfo {volume} {6}},\ \bibinfo {pages} {043161} (\bibinfo {year} {2024})}\BibitemShut {NoStop}%
\bibitem [{\citenamefont {Lasner}\ \emph {et~al.}(2024)\citenamefont {Lasner}, \citenamefont {Frenett}, \citenamefont {Sawaoka}, \citenamefont {Anderegg}, \citenamefont {Augenbraun}, \citenamefont {Lampson}, \citenamefont {Li}, \citenamefont {Lunstad}, \citenamefont {Mango}, \citenamefont {Nasir}, \citenamefont {Ono}, \citenamefont {Sakamoto},\ and\ \citenamefont {Doyle}}]{Lasner2024}%
  \BibitemOpen
  \bibfield  {author} {\bibinfo {author} {\bibfnamefont {Z.~D.}\ \bibnamefont {Lasner}}, \bibinfo {author} {\bibfnamefont {A.}~\bibnamefont {Frenett}}, \bibinfo {author} {\bibfnamefont {H.}~\bibnamefont {Sawaoka}}, \bibinfo {author} {\bibfnamefont {L.}~\bibnamefont {Anderegg}}, \bibinfo {author} {\bibfnamefont {B.}~\bibnamefont {Augenbraun}}, \bibinfo {author} {\bibfnamefont {H.}~\bibnamefont {Lampson}}, \bibinfo {author} {\bibfnamefont {M.}~\bibnamefont {Li}}, \bibinfo {author} {\bibfnamefont {A.}~\bibnamefont {Lunstad}}, \bibinfo {author} {\bibfnamefont {J.}~\bibnamefont {Mango}}, \bibinfo {author} {\bibfnamefont {A.}~\bibnamefont {Nasir}}, \bibinfo {author} {\bibfnamefont {T.}~\bibnamefont {Ono}}, \bibinfo {author} {\bibfnamefont {T.}~\bibnamefont {Sakamoto}},\ and\ \bibinfo {author} {\bibfnamefont {J.~M.}\ \bibnamefont {Doyle}},\ }\href {https://arxiv.org/abs/2409.04948} {\bibinfo {title} {Magneto-optical trapping of a heavy polyatomic molecule for precision measurement}} (\bibinfo {year} {2024}),\
  \Eprint {https://arxiv.org/abs/2409.04948} {arXiv:2409.04948 [physics.atom-ph]} \BibitemShut {NoStop}%
\bibitem [{\citenamefont {Kozyryev}\ and\ \citenamefont {Hutzler}(2017)}]{Kozyryev2017}%
  \BibitemOpen
  \bibfield  {author} {\bibinfo {author} {\bibfnamefont {I.}~\bibnamefont {Kozyryev}}\ and\ \bibinfo {author} {\bibfnamefont {N.~R.}\ \bibnamefont {Hutzler}},\ }\bibfield  {title} {\bibinfo {title} {Precision measurement of time-reversal symmetry violation with laser-cooled polyatomic molecules},\ }\href {https://doi.org/10.1103/PhysRevLett.119.133002} {\bibfield  {journal} {\bibinfo  {journal} {Phys. Rev. Lett.}\ }\textbf {\bibinfo {volume} {119}},\ \bibinfo {pages} {133002} (\bibinfo {year} {2017})}\BibitemShut {NoStop}%
\bibitem [{\citenamefont {Fitch}\ \emph {et~al.}(2021)\citenamefont {Fitch}, \citenamefont {Lim}, \citenamefont {Hinds}, \citenamefont {Sauer},\ and\ \citenamefont {Tarbutt}}]{Fitch2021methods}%
  \BibitemOpen
  \bibfield  {author} {\bibinfo {author} {\bibfnamefont {N.~J.}\ \bibnamefont {Fitch}}, \bibinfo {author} {\bibfnamefont {J.}~\bibnamefont {Lim}}, \bibinfo {author} {\bibfnamefont {E.~A.}\ \bibnamefont {Hinds}}, \bibinfo {author} {\bibfnamefont {B.~E.}\ \bibnamefont {Sauer}},\ and\ \bibinfo {author} {\bibfnamefont {M.~R.}\ \bibnamefont {Tarbutt}},\ }\bibfield  {title} {\bibinfo {title} {{Methods for measuring the electron's electric dipole moment using ultracold YbF molecules}},\ }\href {https://doi.org/10.1088/2058-9565/abc931} {\bibfield  {journal} {\bibinfo  {journal} {Quantum Science and Technology}\ }\textbf {\bibinfo {volume} {6}},\ \bibinfo {pages} {14006} (\bibinfo {year} {2021})}\BibitemShut {NoStop}%
\bibitem [{\citenamefont {Anderegg}\ \emph {et~al.}(2023)\citenamefont {Anderegg}, \citenamefont {Vilas}, \citenamefont {Hallas}, \citenamefont {Robichaud}, \citenamefont {Jadbabaie}, \citenamefont {Doyle},\ and\ \citenamefont {Hutzler}}]{Anderegg2024}%
  \BibitemOpen
  \bibfield  {author} {\bibinfo {author} {\bibfnamefont {L.}~\bibnamefont {Anderegg}}, \bibinfo {author} {\bibfnamefont {N.~B.}\ \bibnamefont {Vilas}}, \bibinfo {author} {\bibfnamefont {C.}~\bibnamefont {Hallas}}, \bibinfo {author} {\bibfnamefont {P.}~\bibnamefont {Robichaud}}, \bibinfo {author} {\bibfnamefont {A.}~\bibnamefont {Jadbabaie}}, \bibinfo {author} {\bibfnamefont {J.~M.}\ \bibnamefont {Doyle}},\ and\ \bibinfo {author} {\bibfnamefont {N.~R.}\ \bibnamefont {Hutzler}},\ }\bibfield  {title} {\bibinfo {title} {Quantum control of trapped polyatomic molecules for eedm searches},\ }\href {https://doi.org/10.1126/science.adg8155} {\bibfield  {journal} {\bibinfo  {journal} {Science}\ }\textbf {\bibinfo {volume} {382}},\ \bibinfo {pages} {665} (\bibinfo {year} {2023})}\BibitemShut {NoStop}%
\bibitem [{\citenamefont {Fitch}\ and\ \citenamefont {Tarbutt}(2021)}]{Fitch2021}%
  \BibitemOpen
  \bibfield  {author} {\bibinfo {author} {\bibfnamefont {N.}~\bibnamefont {Fitch}}\ and\ \bibinfo {author} {\bibfnamefont {M.}~\bibnamefont {Tarbutt}},\ }\bibfield  {title} {\bibinfo {title} {{L}aser-cooled molecules},\ }\href {https://doi.org/https://doi.org/10.1016/bs.aamop.2021.04.003} {\bibfield  {journal} {\bibinfo  {journal} {Advances In Atomic, Molecular, and Optical Physics}\ }\textbf {\bibinfo {volume} {70}},\ \bibinfo {pages} {157} (\bibinfo {year} {2021})}\BibitemShut {NoStop}%
\bibitem [{\citenamefont {Bouchiat}\ and\ \citenamefont {Bouchiat}(1997)}]{Bouchiat1997}%
  \BibitemOpen
  \bibfield  {author} {\bibinfo {author} {\bibfnamefont {M.-A.}\ \bibnamefont {Bouchiat}}\ and\ \bibinfo {author} {\bibfnamefont {C.}~\bibnamefont {Bouchiat}},\ }\bibfield  {title} {\bibinfo {title} {Parity violation in atoms},\ }\href {https://doi.org/10.1088/0034-4885/60/11/004} {\bibfield  {journal} {\bibinfo  {journal} {Reports on Progress in Physics}\ }\textbf {\bibinfo {volume} {60}},\ \bibinfo {pages} {1351} (\bibinfo {year} {1997})}\BibitemShut {NoStop}%
\bibitem [{\citenamefont {{The Jefferson Lab PVDIS Collaboration}}(2014)}]{Wang2014}%
  \BibitemOpen
  \bibfield  {author} {\bibinfo {author} {\bibnamefont {{The Jefferson Lab PVDIS Collaboration}}},\ }\bibfield  {title} {\bibinfo {title} {Measurement of parity violation in electron--quark scattering},\ }\href {https://doi.org/10.1038/nature12964} {\bibfield  {journal} {\bibinfo  {journal} {Nature}\ }\textbf {\bibinfo {volume} {506}},\ \bibinfo {pages} {67} (\bibinfo {year} {2014})}\BibitemShut {NoStop}%
\bibitem [{\citenamefont {Altuntas}\ \emph {et~al.}(2018)\citenamefont {Altuntas}, \citenamefont {Ammon}, \citenamefont {Cahn},\ and\ \citenamefont {DeMille}}]{Altuntas2018}%
  \BibitemOpen
  \bibfield  {author} {\bibinfo {author} {\bibfnamefont {E.}~\bibnamefont {Altuntas}}, \bibinfo {author} {\bibfnamefont {J.}~\bibnamefont {Ammon}}, \bibinfo {author} {\bibfnamefont {S.~B.}\ \bibnamefont {Cahn}},\ and\ \bibinfo {author} {\bibfnamefont {D.}~\bibnamefont {DeMille}},\ }\bibfield  {title} {\bibinfo {title} {{Demonstration of a Sensitive Method to Measure Nuclear-Spin-Dependent Parity Violation}},\ }\href {https://doi.org/10.1103/PhysRevLett.120.142501} {\bibfield  {journal} {\bibinfo  {journal} {Phys. Rev. Lett.}\ }\textbf {\bibinfo {volume} {120}},\ \bibinfo {pages} {142501} (\bibinfo {year} {2018})}\BibitemShut {NoStop}%
\bibitem [{\citenamefont {Antypas}\ \emph {et~al.}(2019)\citenamefont {Antypas}, \citenamefont {Fabricant}, \citenamefont {Stalnaker}, \citenamefont {Tsigutkin}, \citenamefont {Flambaum},\ and\ \citenamefont {Budker}}]{Antypas2019}%
  \BibitemOpen
  \bibfield  {author} {\bibinfo {author} {\bibfnamefont {D.}~\bibnamefont {Antypas}}, \bibinfo {author} {\bibfnamefont {A.}~\bibnamefont {Fabricant}}, \bibinfo {author} {\bibfnamefont {J.~E.}\ \bibnamefont {Stalnaker}}, \bibinfo {author} {\bibfnamefont {K.}~\bibnamefont {Tsigutkin}}, \bibinfo {author} {\bibfnamefont {V.~V.}\ \bibnamefont {Flambaum}},\ and\ \bibinfo {author} {\bibfnamefont {D.}~\bibnamefont {Budker}},\ }\bibfield  {title} {\bibinfo {title} {Isotopic variation of parity violation in atomic ytterbium},\ }\href {https://doi.org/10.1038/s41567-018-0312-8} {\bibfield  {journal} {\bibinfo  {journal} {Nature Physics}\ }\textbf {\bibinfo {volume} {15}},\ \bibinfo {pages} {120} (\bibinfo {year} {2019})}\BibitemShut {NoStop}%
\bibitem [{\citenamefont {Wood}\ \emph {et~al.}(1997)\citenamefont {Wood}, \citenamefont {Bennett}, \citenamefont {Cho}, \citenamefont {Masterson}, \citenamefont {Roberts}, \citenamefont {Tanner},\ and\ \citenamefont {Wieman}}]{Wood1997}%
  \BibitemOpen
  \bibfield  {author} {\bibinfo {author} {\bibfnamefont {C.~S.}\ \bibnamefont {Wood}}, \bibinfo {author} {\bibfnamefont {S.~C.}\ \bibnamefont {Bennett}}, \bibinfo {author} {\bibfnamefont {D.}~\bibnamefont {Cho}}, \bibinfo {author} {\bibfnamefont {B.~P.}\ \bibnamefont {Masterson}}, \bibinfo {author} {\bibfnamefont {J.~L.}\ \bibnamefont {Roberts}}, \bibinfo {author} {\bibfnamefont {C.~E.}\ \bibnamefont {Tanner}},\ and\ \bibinfo {author} {\bibfnamefont {C.~E.}\ \bibnamefont {Wieman}},\ }\bibfield  {title} {\bibinfo {title} {{Measurement of Parity Nonconservation and an Anapole Moment in Cesium}},\ }\href {https://doi.org/10.1126/science.275.5307.1759} {\bibfield  {journal} {\bibinfo  {journal} {Science}\ }\textbf {\bibinfo {volume} {275}},\ \bibinfo {pages} {1759} (\bibinfo {year} {1997})}\BibitemShut {NoStop}%
\bibitem [{\citenamefont {Haxton}\ \emph {et~al.}(2002)\citenamefont {Haxton}, \citenamefont {Liu},\ and\ \citenamefont {Ramsey-Musolf}}]{Haxton2002}%
  \BibitemOpen
  \bibfield  {author} {\bibinfo {author} {\bibfnamefont {W.~C.}\ \bibnamefont {Haxton}}, \bibinfo {author} {\bibfnamefont {C.-P.}\ \bibnamefont {Liu}},\ and\ \bibinfo {author} {\bibfnamefont {M.~J.}\ \bibnamefont {Ramsey-Musolf}},\ }\bibfield  {title} {\bibinfo {title} {Nuclear anapole moments},\ }\href {https://doi.org/10.1103/PhysRevC.65.045502} {\bibfield  {journal} {\bibinfo  {journal} {Phys. Rev. C}\ }\textbf {\bibinfo {volume} {65}},\ \bibinfo {pages} {045502} (\bibinfo {year} {2002})}\BibitemShut {NoStop}%
\bibitem [{\citenamefont {Johnson}\ \emph {et~al.}(2003)\citenamefont {Johnson}, \citenamefont {Safronova},\ and\ \citenamefont {Safronova}}]{Johnson2003}%
  \BibitemOpen
  \bibfield  {author} {\bibinfo {author} {\bibfnamefont {W.~R.}\ \bibnamefont {Johnson}}, \bibinfo {author} {\bibfnamefont {M.~S.}\ \bibnamefont {Safronova}},\ and\ \bibinfo {author} {\bibfnamefont {U.~I.}\ \bibnamefont {Safronova}},\ }\bibfield  {title} {\bibinfo {title} {Combined effect of coherent z exchange and the hyperfine interaction in the atomic parity-nonconserving interaction},\ }\href {https://doi.org/10.1103/PhysRevA.67.062106} {\bibfield  {journal} {\bibinfo  {journal} {Phys. Rev. A}\ }\textbf {\bibinfo {volume} {67}},\ \bibinfo {pages} {062106} (\bibinfo {year} {2003})}\BibitemShut {NoStop}%
\bibitem [{\citenamefont {Kozlov}\ and\ \citenamefont {Labzowsky}(1995)}]{Kozlov1995}%
  \BibitemOpen
  \bibfield  {author} {\bibinfo {author} {\bibfnamefont {M.~G.}\ \bibnamefont {Kozlov}}\ and\ \bibinfo {author} {\bibfnamefont {L.~N.}\ \bibnamefont {Labzowsky}},\ }\bibfield  {title} {\bibinfo {title} {Parity violation effects in diatomics},\ }\href {https://doi.org/10.1088/0953-4075/28/10/008} {\bibfield  {journal} {\bibinfo  {journal} {Journal of Physics B: Atomic, Molecular and Optical Physics}\ }\textbf {\bibinfo {volume} {28}},\ \bibinfo {pages} {1933} (\bibinfo {year} {1995})}\BibitemShut {NoStop}%
\bibitem [{\citenamefont {DeMille}\ \emph {et~al.}(2008)\citenamefont {DeMille}, \citenamefont {Cahn}, \citenamefont {Murphree}, \citenamefont {Rahmlow},\ and\ \citenamefont {Kozlov}}]{DeMille2008}%
  \BibitemOpen
  \bibfield  {author} {\bibinfo {author} {\bibfnamefont {D.}~\bibnamefont {DeMille}}, \bibinfo {author} {\bibfnamefont {S.~B.}\ \bibnamefont {Cahn}}, \bibinfo {author} {\bibfnamefont {D.}~\bibnamefont {Murphree}}, \bibinfo {author} {\bibfnamefont {D.~A.}\ \bibnamefont {Rahmlow}},\ and\ \bibinfo {author} {\bibfnamefont {M.~G.}\ \bibnamefont {Kozlov}},\ }\bibfield  {title} {\bibinfo {title} {Using molecules to measure nuclear spin-dependent parity violation},\ }\href {https://doi.org/10.1103/PhysRevLett.100.023003} {\bibfield  {journal} {\bibinfo  {journal} {Phys. Rev. Lett.}\ }\textbf {\bibinfo {volume} {100}},\ \bibinfo {pages} {023003} (\bibinfo {year} {2008})}\BibitemShut {NoStop}%
\bibitem [{\citenamefont {Altunta\ifmmode~\mbox{\c{s}}\else \c{s}\fi{}}\ \emph {et~al.}(2018)\citenamefont {Altunta\ifmmode~\mbox{\c{s}}\else \c{s}\fi{}}, \citenamefont {Ammon}, \citenamefont {Cahn},\ and\ \citenamefont {DeMille}}]{AltuntasPRA}%
  \BibitemOpen
  \bibfield  {author} {\bibinfo {author} {\bibfnamefont {E.}~\bibnamefont {Altunta\ifmmode~\mbox{\c{s}}\else \c{s}\fi{}}}, \bibinfo {author} {\bibfnamefont {J.}~\bibnamefont {Ammon}}, \bibinfo {author} {\bibfnamefont {S.~B.}\ \bibnamefont {Cahn}},\ and\ \bibinfo {author} {\bibfnamefont {D.}~\bibnamefont {DeMille}},\ }\bibfield  {title} {\bibinfo {title} {Measuring nuclear-spin-dependent parity violation with molecules: Experimental methods and analysis of systematic errors},\ }\href {https://doi.org/10.1103/PhysRevA.97.042101} {\bibfield  {journal} {\bibinfo  {journal} {Phys. Rev. A}\ }\textbf {\bibinfo {volume} {97}},\ \bibinfo {pages} {042101} (\bibinfo {year} {2018})}\BibitemShut {NoStop}%
\bibitem [{\citenamefont {Dzuba}\ \emph {et~al.}(2017)\citenamefont {Dzuba}, \citenamefont {Flambaum},\ and\ \citenamefont {Stadnik}}]{Dzuba2017}%
  \BibitemOpen
  \bibfield  {author} {\bibinfo {author} {\bibfnamefont {V.~A.}\ \bibnamefont {Dzuba}}, \bibinfo {author} {\bibfnamefont {V.~V.}\ \bibnamefont {Flambaum}},\ and\ \bibinfo {author} {\bibfnamefont {Y.~V.}\ \bibnamefont {Stadnik}},\ }\bibfield  {title} {\bibinfo {title} {{Probing Low-Mass Vector Bosons with Parity Nonconservation and Nuclear Anapole Moment Measurements in Atoms and Molecules}},\ }\href {https://doi.org/10.1103/PhysRevLett.119.223201} {\bibfield  {journal} {\bibinfo  {journal} {Phys. Rev. Lett.}\ }\textbf {\bibinfo {volume} {119}},\ \bibinfo {pages} {223201} (\bibinfo {year} {2017})}\BibitemShut {NoStop}%
\bibitem [{\citenamefont {Gaul}\ \emph {et~al.}(2020)\citenamefont {Gaul}, \citenamefont {Kozlov}, \citenamefont {Isaev},\ and\ \citenamefont {Berger}}]{Gaul2020}%
  \BibitemOpen
  \bibfield  {author} {\bibinfo {author} {\bibfnamefont {K.}~\bibnamefont {Gaul}}, \bibinfo {author} {\bibfnamefont {M.~G.}\ \bibnamefont {Kozlov}}, \bibinfo {author} {\bibfnamefont {T.~A.}\ \bibnamefont {Isaev}},\ and\ \bibinfo {author} {\bibfnamefont {R.}~\bibnamefont {Berger}},\ }\bibfield  {title} {\bibinfo {title} {Chiral molecules as sensitive probes for direct detection of $\mathcal{P}$-odd cosmic fields},\ }\href {https://doi.org/10.1103/PhysRevLett.125.123004} {\bibfield  {journal} {\bibinfo  {journal} {Phys. Rev. Lett.}\ }\textbf {\bibinfo {volume} {125}},\ \bibinfo {pages} {123004} (\bibinfo {year} {2020})}\BibitemShut {NoStop}%
\bibitem [{\citenamefont {Flambaum}\ and\ \citenamefont {Khriplovich}(1985)}]{Flambaum1985}%
  \BibitemOpen
  \bibfield  {author} {\bibinfo {author} {\bibfnamefont {V.~V.}\ \bibnamefont {Flambaum}}\ and\ \bibinfo {author} {\bibfnamefont {I.~B.}\ \bibnamefont {Khriplovich}},\ }\bibfield  {title} {\bibinfo {title} {On the enhancement of parity nonconserving effects in diatomic molecules},\ }\href {https://doi.org/https://doi.org/10.1016/0375-9601(85)90756-X} {\bibfield  {journal} {\bibinfo  {journal} {Physics Letters A}\ }\textbf {\bibinfo {volume} {110}},\ \bibinfo {pages} {121} (\bibinfo {year} {1985})}\BibitemShut {NoStop}%
\bibitem [{\citenamefont {Kogel}\ \emph {et~al.}(2021)\citenamefont {Kogel}, \citenamefont {Rockenhäuser}, \citenamefont {Albrecht},\ and\ \citenamefont {Langen}}]{Kogel2021}%
  \BibitemOpen
  \bibfield  {author} {\bibinfo {author} {\bibfnamefont {F.}~\bibnamefont {Kogel}}, \bibinfo {author} {\bibfnamefont {M.}~\bibnamefont {Rockenhäuser}}, \bibinfo {author} {\bibfnamefont {R.}~\bibnamefont {Albrecht}},\ and\ \bibinfo {author} {\bibfnamefont {T.}~\bibnamefont {Langen}},\ }\bibfield  {title} {\bibinfo {title} {A laser cooling scheme for precision measurements using fermionic barium monofluoride {(137Ba19F)} molecules},\ }\href {https://doi.org/10.1088/1367-2630/ac1df2} {\bibfield  {journal} {\bibinfo  {journal} {New Journal of Physics}\ }\textbf {\bibinfo {volume} {23}},\ \bibinfo {pages} {095003} (\bibinfo {year} {2021})}\BibitemShut {NoStop}%
\bibitem [{\citenamefont {Norrgard}\ \emph {et~al.}(2019)\citenamefont {Norrgard}, \citenamefont {Barker}, \citenamefont {Eckel}, \citenamefont {Fedchak}, \citenamefont {Klimov},\ and\ \citenamefont {Scherschligt}}]{Norrgard2019}%
  \BibitemOpen
  \bibfield  {author} {\bibinfo {author} {\bibfnamefont {E.~B.}\ \bibnamefont {Norrgard}}, \bibinfo {author} {\bibfnamefont {D.~S.}\ \bibnamefont {Barker}}, \bibinfo {author} {\bibfnamefont {S.}~\bibnamefont {Eckel}}, \bibinfo {author} {\bibfnamefont {J.~A.}\ \bibnamefont {Fedchak}}, \bibinfo {author} {\bibfnamefont {N.~N.}\ \bibnamefont {Klimov}},\ and\ \bibinfo {author} {\bibfnamefont {J.}~\bibnamefont {Scherschligt}},\ }\bibfield  {title} {\bibinfo {title} {{Nuclear-spin dependent parity violation in optically trapped polyatomic molecules}},\ }\href {https://doi.org/10.1038/s42005-019-0181-1} {\bibfield  {journal} {\bibinfo  {journal} {Communications Physics}\ }\textbf {\bibinfo {volume} {2}},\ \bibinfo {pages} {77} (\bibinfo {year} {2019})}\BibitemShut {NoStop}%
\bibitem [{\citenamefont {Karthein}\ \emph {et~al.}(2024)\citenamefont {Karthein}, \citenamefont {Udrescu}, \citenamefont {Moroch}, \citenamefont {Belosevic}, \citenamefont {Blaum}, \citenamefont {Borschevsky}, \citenamefont {Chamorro}, \citenamefont {DeMille}, \citenamefont {Dilling}, \citenamefont {Garcia~Ruiz}, \citenamefont {Hutzler}, \citenamefont {Pa\ifmmode~\check{s}\else \v{s}\fi{}teka},\ and\ \citenamefont {Ringle}}]{Karthein2024}%
  \BibitemOpen
  \bibfield  {author} {\bibinfo {author} {\bibfnamefont {J.}~\bibnamefont {Karthein}}, \bibinfo {author} {\bibfnamefont {S.~M.}\ \bibnamefont {Udrescu}}, \bibinfo {author} {\bibfnamefont {S.~B.}\ \bibnamefont {Moroch}}, \bibinfo {author} {\bibfnamefont {I.}~\bibnamefont {Belosevic}}, \bibinfo {author} {\bibfnamefont {K.}~\bibnamefont {Blaum}}, \bibinfo {author} {\bibfnamefont {A.}~\bibnamefont {Borschevsky}}, \bibinfo {author} {\bibfnamefont {Y.}~\bibnamefont {Chamorro}}, \bibinfo {author} {\bibfnamefont {D.}~\bibnamefont {DeMille}}, \bibinfo {author} {\bibfnamefont {J.}~\bibnamefont {Dilling}}, \bibinfo {author} {\bibfnamefont {R.~F.}\ \bibnamefont {Garcia~Ruiz}}, \bibinfo {author} {\bibfnamefont {N.~R.}\ \bibnamefont {Hutzler}}, \bibinfo {author} {\bibfnamefont {L.~F.}\ \bibnamefont {Pa\ifmmode~\check{s}\else \v{s}\fi{}teka}},\ and\ \bibinfo {author} {\bibfnamefont {R.}~\bibnamefont {Ringle}},\ }\bibfield  {title} {\bibinfo {title} {Electroweak nuclear properties from single molecular ions in a penning
  trap},\ }\href {https://doi.org/10.1103/PhysRevLett.133.033003} {\bibfield  {journal} {\bibinfo  {journal} {Phys. Rev. Lett.}\ }\textbf {\bibinfo {volume} {133}},\ \bibinfo {pages} {033003} (\bibinfo {year} {2024})}\BibitemShut {NoStop}%
\bibitem [{\citenamefont {Kogel}\ \emph {et~al.}(2025{\natexlab{a}})\citenamefont {Kogel}, \citenamefont {Garg}, \citenamefont {Rockenh{\"{a}}user}, \citenamefont {Morales-Ram{\'{i}}rez},\ and\ \citenamefont {Langen}}]{Kogel2024isotope}%
  \BibitemOpen
  \bibfield  {author} {\bibinfo {author} {\bibfnamefont {F.}~\bibnamefont {Kogel}}, \bibinfo {author} {\bibfnamefont {T.}~\bibnamefont {Garg}}, \bibinfo {author} {\bibfnamefont {M.}~\bibnamefont {Rockenh{\"{a}}user}}, \bibinfo {author} {\bibfnamefont {S.~A.}\ \bibnamefont {Morales-Ram{\'{i}}rez}},\ and\ \bibinfo {author} {\bibfnamefont {T.}~\bibnamefont {Langen}},\ }\bibfield  {title} {\bibinfo {title} {{Isotopologue-selective laser cooling of molecules}},\ }\href {https://doi.org/10.1088/1367-2630/ada3f0} {\bibfield  {journal} {\bibinfo  {journal} {New Journal of Physics}\ }\textbf {\bibinfo {volume} {27}},\ \bibinfo {pages} {13001} (\bibinfo {year} {2025}{\natexlab{a}})}\BibitemShut {NoStop}%
\bibitem [{\citenamefont {Albrecht}\ \emph {et~al.}(2020)\citenamefont {Albrecht}, \citenamefont {Scharwaechter}, \citenamefont {Sixt}, \citenamefont {Hofer},\ and\ \citenamefont {Langen}}]{Albrecht2020}%
  \BibitemOpen
  \bibfield  {author} {\bibinfo {author} {\bibfnamefont {R.}~\bibnamefont {Albrecht}}, \bibinfo {author} {\bibfnamefont {M.}~\bibnamefont {Scharwaechter}}, \bibinfo {author} {\bibfnamefont {T.}~\bibnamefont {Sixt}}, \bibinfo {author} {\bibfnamefont {L.}~\bibnamefont {Hofer}},\ and\ \bibinfo {author} {\bibfnamefont {T.}~\bibnamefont {Langen}},\ }\bibfield  {title} {\bibinfo {title} {Buffer-gas cooling, high-resolution spectroscopy, and optical cycling of barium monofluoride molecules},\ }\href {https://doi.org/10.1103/PhysRevA.101.013413} {\bibfield  {journal} {\bibinfo  {journal} {Phys. Rev. A}\ }\textbf {\bibinfo {volume} {101}},\ \bibinfo {pages} {013413} (\bibinfo {year} {2020})}\BibitemShut {NoStop}%
\bibitem [{\citenamefont {Rockenh\"auser}\ \emph {et~al.}(2023)\citenamefont {Rockenh\"auser}, \citenamefont {Kogel}, \citenamefont {Pultinevicius},\ and\ \citenamefont {Langen}}]{Rockenhaeuser2023}%
  \BibitemOpen
  \bibfield  {author} {\bibinfo {author} {\bibfnamefont {M.}~\bibnamefont {Rockenh\"auser}}, \bibinfo {author} {\bibfnamefont {F.}~\bibnamefont {Kogel}}, \bibinfo {author} {\bibfnamefont {E.}~\bibnamefont {Pultinevicius}},\ and\ \bibinfo {author} {\bibfnamefont {T.}~\bibnamefont {Langen}},\ }\bibfield  {title} {\bibinfo {title} {Absorption spectroscopy for laser cooling and high-fidelity detection of barium monofluoride molecules},\ }\href {https://doi.org/10.1103/PhysRevA.108.062812} {\bibfield  {journal} {\bibinfo  {journal} {Phys. Rev. A}\ }\textbf {\bibinfo {volume} {108}},\ \bibinfo {pages} {062812} (\bibinfo {year} {2023})}\BibitemShut {NoStop}%
\bibitem [{\citenamefont {Aggarwal}\ \emph {et~al.}(2018)\citenamefont {Aggarwal}, \citenamefont {Bethlem}, \citenamefont {Borschevsky}, \citenamefont {Denis}, \citenamefont {Esajas}, \citenamefont {Haase}, \citenamefont {Hao}, \citenamefont {Hoekstra}, \citenamefont {Jungmann}, \citenamefont {Meijknecht}, \citenamefont {Mooij}, \citenamefont {Timmermans}, \citenamefont {Ubachs}, \citenamefont {Willmann},\ and\ \citenamefont {Zapara}}]{Aggarwal2018}%
  \BibitemOpen
  \bibfield  {author} {\bibinfo {author} {\bibfnamefont {P.}~\bibnamefont {Aggarwal}}, \bibinfo {author} {\bibfnamefont {H.~L.}\ \bibnamefont {Bethlem}}, \bibinfo {author} {\bibfnamefont {A.}~\bibnamefont {Borschevsky}}, \bibinfo {author} {\bibfnamefont {M.}~\bibnamefont {Denis}}, \bibinfo {author} {\bibfnamefont {K.}~\bibnamefont {Esajas}}, \bibinfo {author} {\bibfnamefont {P.~A.~B.}\ \bibnamefont {Haase}}, \bibinfo {author} {\bibfnamefont {Y.}~\bibnamefont {Hao}}, \bibinfo {author} {\bibfnamefont {S.}~\bibnamefont {Hoekstra}}, \bibinfo {author} {\bibfnamefont {K.}~\bibnamefont {Jungmann}}, \bibinfo {author} {\bibfnamefont {T.~B.}\ \bibnamefont {Meijknecht}}, \bibinfo {author} {\bibfnamefont {M.~C.}\ \bibnamefont {Mooij}}, \bibinfo {author} {\bibfnamefont {R.~G.~E.}\ \bibnamefont {Timmermans}}, \bibinfo {author} {\bibfnamefont {W.}~\bibnamefont {Ubachs}}, \bibinfo {author} {\bibfnamefont {L.}~\bibnamefont {Willmann}},\ and\ \bibinfo {author} {\bibfnamefont {A.}~\bibnamefont {Zapara}},\ }\bibfield  {title}
  {\bibinfo {title} {{Measuring the electric dipole moment of the electron in BaF}},\ }\href {https://doi.org/10.1140/epjd/e2018-90192-9} {\bibfield  {journal} {\bibinfo  {journal} {Eur. Phys. J. D}\ }\textbf {\bibinfo {volume} {72}},\ \bibinfo {pages} {197} (\bibinfo {year} {2018})}\BibitemShut {NoStop}%
\bibitem [{\citenamefont {Kogel}\ \emph {et~al.}(2025{\natexlab{b}})\citenamefont {Kogel}, \citenamefont {Garg}, \citenamefont {Rockenhäuser}, \citenamefont {Morales-Ramírez},\ and\ \citenamefont {Langen}}]{Kogel2024serrodynes}%
  \BibitemOpen
  \bibfield  {author} {\bibinfo {author} {\bibfnamefont {F.}~\bibnamefont {Kogel}}, \bibinfo {author} {\bibfnamefont {T.}~\bibnamefont {Garg}}, \bibinfo {author} {\bibfnamefont {M.}~\bibnamefont {Rockenhäuser}}, \bibinfo {author} {\bibfnamefont {S.~A.}\ \bibnamefont {Morales-Ramírez}},\ and\ \bibinfo {author} {\bibfnamefont {T.}~\bibnamefont {Langen}},\ }\href {https://arxiv.org/abs/2501.10725} {\bibinfo {title} {Molecular laser cooling using serrodynes: {I}mplementation, characterization and prospects}} (\bibinfo {year} {2025}{\natexlab{b}}),\ \Eprint {https://arxiv.org/abs/2501.10725} {arXiv:2501.10725 [physics.atom-ph]} \BibitemShut {NoStop}%
\bibitem [{\citenamefont {Zeng}\ \emph {et~al.}(2024)\citenamefont {Zeng}, \citenamefont {Deng}, \citenamefont {Yang},\ and\ \citenamefont {Yan}}]{Zeng2024}%
  \BibitemOpen
  \bibfield  {author} {\bibinfo {author} {\bibfnamefont {Z.}~\bibnamefont {Zeng}}, \bibinfo {author} {\bibfnamefont {S.}~\bibnamefont {Deng}}, \bibinfo {author} {\bibfnamefont {S.}~\bibnamefont {Yang}},\ and\ \bibinfo {author} {\bibfnamefont {B.}~\bibnamefont {Yan}},\ }\bibfield  {title} {\bibinfo {title} {Three-dimensional magneto-optical trapping of barium monofluoride},\ }\href {https://doi.org/10.1103/PhysRevLett.133.143404} {\bibfield  {journal} {\bibinfo  {journal} {Phys. Rev. Lett.}\ }\textbf {\bibinfo {volume} {133}},\ \bibinfo {pages} {143404} (\bibinfo {year} {2024})}\BibitemShut {NoStop}%
\bibitem [{\citenamefont {{Kogel \textit{et al.}}}(2025)}]{Kogel2024fermispectroscopy}%
  \BibitemOpen
  \bibfield  {author} {\bibinfo {author} {\bibfnamefont {F.}~\bibnamefont {{Kogel \textit{et al.}}}},\ }\href@noop {} {\bibinfo {title} {High-resolution spectroscopy of barium monofluoride: Observation of odd–even staggering of nuclear charge radii in a molecule (\textit{in preparation})}} (\bibinfo {year} {2025})\BibitemShut {NoStop}%
\bibitem [{\citenamefont {Steimle}\ \emph {et~al.}(2011)\citenamefont {Steimle}, \citenamefont {Frey}, \citenamefont {Le}, \citenamefont {DeMille}, \citenamefont {Rahmlow},\ and\ \citenamefont {Linton}}]{Steimle2011}%
  \BibitemOpen
  \bibfield  {author} {\bibinfo {author} {\bibfnamefont {T.~C.}\ \bibnamefont {Steimle}}, \bibinfo {author} {\bibfnamefont {S.}~\bibnamefont {Frey}}, \bibinfo {author} {\bibfnamefont {A.}~\bibnamefont {Le}}, \bibinfo {author} {\bibfnamefont {D.}~\bibnamefont {DeMille}}, \bibinfo {author} {\bibfnamefont {D.~A.}\ \bibnamefont {Rahmlow}},\ and\ \bibinfo {author} {\bibfnamefont {C.}~\bibnamefont {Linton}},\ }\bibfield  {title} {\bibinfo {title} {Molecular-beam optical {Stark} and {Zeeman} study of the {$A$} ${}^{2}\ensuremath{\Pi}$--{$X$} ${}^{2}{\ensuremath{\Sigma}}^{+}$ (0,0) band system of {BaF}},\ }\href {https://doi.org/10.1103/PhysRevA.84.012508} {\bibfield  {journal} {\bibinfo  {journal} {Phys. Rev. A}\ }\textbf {\bibinfo {volume} {84}},\ \bibinfo {pages} {012508} (\bibinfo {year} {2011})}\BibitemShut {NoStop}%
\bibitem [{\citenamefont {Pultinevicius}\ \emph {et~al.}(2023)\citenamefont {Pultinevicius}, \citenamefont {Rockenhäuser}, \citenamefont {Kogel}, \citenamefont {Groß}, \citenamefont {Garg}, \citenamefont {Prochnow},\ and\ \citenamefont {Langen}}]{Pultinevicius2023}%
  \BibitemOpen
  \bibfield  {author} {\bibinfo {author} {\bibfnamefont {E.}~\bibnamefont {Pultinevicius}}, \bibinfo {author} {\bibfnamefont {M.}~\bibnamefont {Rockenhäuser}}, \bibinfo {author} {\bibfnamefont {F.}~\bibnamefont {Kogel}}, \bibinfo {author} {\bibfnamefont {P.}~\bibnamefont {Groß}}, \bibinfo {author} {\bibfnamefont {T.}~\bibnamefont {Garg}}, \bibinfo {author} {\bibfnamefont {O.~E.}\ \bibnamefont {Prochnow}},\ and\ \bibinfo {author} {\bibfnamefont {T.}~\bibnamefont {Langen}},\ }\bibfield  {title} {\bibinfo {title} {A scalable scanning transfer cavity laser stabilization scheme based on the {Red Pitaya STEMlab} platform},\ }\href {https://doi.org/10.1063/5.0169021} {\bibfield  {journal} {\bibinfo  {journal} {Review of Scientific Instruments}\ }\textbf {\bibinfo {volume} {94}},\ \bibinfo {pages} {103004} (\bibinfo {year} {2023})}\BibitemShut {NoStop}%
\bibitem [{\citenamefont {Lasner}\ and\ \citenamefont {DeMille}(2018)}]{Lasner2018}%
  \BibitemOpen
  \bibfield  {author} {\bibinfo {author} {\bibfnamefont {Z.}~\bibnamefont {Lasner}}\ and\ \bibinfo {author} {\bibfnamefont {D.}~\bibnamefont {DeMille}},\ }\bibfield  {title} {\bibinfo {title} {Statistical sensitivity of phase measurements via laser-induced fluorescence with optical cycling detection},\ }\href {https://doi.org/10.1103/PhysRevA.98.053823} {\bibfield  {journal} {\bibinfo  {journal} {Phys. Rev. A}\ }\textbf {\bibinfo {volume} {98}},\ \bibinfo {pages} {053823} (\bibinfo {year} {2018})}\BibitemShut {NoStop}%
\bibitem [{\citenamefont {Shaw}\ \emph {et~al.}(2021)\citenamefont {Shaw}, \citenamefont {Schnaubelt},\ and\ \citenamefont {McCarron}}]{Shaw2021}%
  \BibitemOpen
  \bibfield  {author} {\bibinfo {author} {\bibfnamefont {J.~C.}\ \bibnamefont {Shaw}}, \bibinfo {author} {\bibfnamefont {J.~C.}\ \bibnamefont {Schnaubelt}},\ and\ \bibinfo {author} {\bibfnamefont {D.~J.}\ \bibnamefont {McCarron}},\ }\bibfield  {title} {\bibinfo {title} {Resonance raman optical cycling for high-fidelity fluorescence detection of molecules},\ }\href {https://doi.org/10.1103/PhysRevResearch.3.L042041} {\bibfield  {journal} {\bibinfo  {journal} {Phys. Rev. Research}\ }\textbf {\bibinfo {volume} {3}},\ \bibinfo {pages} {L042041} (\bibinfo {year} {2021})}\BibitemShut {NoStop}%
\bibitem [{\citenamefont {Mitra}\ \emph {et~al.}(2020)\citenamefont {Mitra}, \citenamefont {Vilas}, \citenamefont {Hallas}, \citenamefont {Anderegg}, \citenamefont {Augenbraun}, \citenamefont {Baum}, \citenamefont {Miller}, \citenamefont {Raval},\ and\ \citenamefont {Doyle}}]{Mitra2020}%
  \BibitemOpen
  \bibfield  {author} {\bibinfo {author} {\bibfnamefont {D.}~\bibnamefont {Mitra}}, \bibinfo {author} {\bibfnamefont {N.~B.}\ \bibnamefont {Vilas}}, \bibinfo {author} {\bibfnamefont {C.}~\bibnamefont {Hallas}}, \bibinfo {author} {\bibfnamefont {L.}~\bibnamefont {Anderegg}}, \bibinfo {author} {\bibfnamefont {B.~L.}\ \bibnamefont {Augenbraun}}, \bibinfo {author} {\bibfnamefont {L.}~\bibnamefont {Baum}}, \bibinfo {author} {\bibfnamefont {C.}~\bibnamefont {Miller}}, \bibinfo {author} {\bibfnamefont {S.}~\bibnamefont {Raval}},\ and\ \bibinfo {author} {\bibfnamefont {J.~M.}\ \bibnamefont {Doyle}},\ }\bibfield  {title} {\bibinfo {title} {Direct laser cooling of a symmetric top molecule},\ }\href {https://doi.org/10.1126/science.abc5357} {\bibfield  {journal} {\bibinfo  {journal} {Science}\ }\textbf {\bibinfo {volume} {369}},\ \bibinfo {pages} {1366} (\bibinfo {year} {2020})}\BibitemShut {NoStop}%
\bibitem [{\citenamefont {Vilas}\ \emph {et~al.}(2022)\citenamefont {Vilas}, \citenamefont {Hallas}, \citenamefont {Anderegg}, \citenamefont {Robichaud}, \citenamefont {Winnicki}, \citenamefont {Mitra},\ and\ \citenamefont {Doyle}}]{Vilas2021}%
  \BibitemOpen
  \bibfield  {author} {\bibinfo {author} {\bibfnamefont {N.~B.}\ \bibnamefont {Vilas}}, \bibinfo {author} {\bibfnamefont {C.}~\bibnamefont {Hallas}}, \bibinfo {author} {\bibfnamefont {L.}~\bibnamefont {Anderegg}}, \bibinfo {author} {\bibfnamefont {P.}~\bibnamefont {Robichaud}}, \bibinfo {author} {\bibfnamefont {A.}~\bibnamefont {Winnicki}}, \bibinfo {author} {\bibfnamefont {D.}~\bibnamefont {Mitra}},\ and\ \bibinfo {author} {\bibfnamefont {J.~M.}\ \bibnamefont {Doyle}},\ }\bibfield  {title} {\bibinfo {title} {Magneto-optical trapping and sub-doppler cooling of a polyatomic molecule},\ }\href {https://doi.org/10.1038/s41586-022-04620-5} {\bibfield  {journal} {\bibinfo  {journal} {Nature}\ }\textbf {\bibinfo {volume} {606}},\ \bibinfo {pages} {70} (\bibinfo {year} {2022})}\BibitemShut {NoStop}%
\bibitem [{\citenamefont {Kudashov}\ \emph {et~al.}(2014)\citenamefont {Kudashov}, \citenamefont {Petrov}, \citenamefont {Skripnikov}, \citenamefont {Mosyagin}, \citenamefont {Isaev}, \citenamefont {Berger},\ and\ \citenamefont {Titov}}]{Kudashov2014}%
  \BibitemOpen
  \bibfield  {author} {\bibinfo {author} {\bibfnamefont {A.~D.}\ \bibnamefont {Kudashov}}, \bibinfo {author} {\bibfnamefont {A.~N.}\ \bibnamefont {Petrov}}, \bibinfo {author} {\bibfnamefont {L.~V.}\ \bibnamefont {Skripnikov}}, \bibinfo {author} {\bibfnamefont {N.~S.}\ \bibnamefont {Mosyagin}}, \bibinfo {author} {\bibfnamefont {T.~A.}\ \bibnamefont {Isaev}}, \bibinfo {author} {\bibfnamefont {R.}~\bibnamefont {Berger}},\ and\ \bibinfo {author} {\bibfnamefont {A.~V.}\ \bibnamefont {Titov}},\ }\bibfield  {title} {\bibinfo {title} {Ab initio study of radium monofluoride (raf) as a candidate to search for parity- and time-and-parity--violation effects},\ }\href {https://doi.org/10.1103/PhysRevA.90.052513} {\bibfield  {journal} {\bibinfo  {journal} {Phys. Rev. A}\ }\textbf {\bibinfo {volume} {90}},\ \bibinfo {pages} {052513} (\bibinfo {year} {2014})}\BibitemShut {NoStop}%
\bibitem [{\citenamefont {Hao}\ \emph {et~al.}(2020)\citenamefont {Hao}, \citenamefont {Navr\'atil}, \citenamefont {Norrgard}, \citenamefont {Ilia\ifmmode~\check{s}\else \v{s}\fi{}}, \citenamefont {Eliav}, \citenamefont {Timmermans}, \citenamefont {Flambaum},\ and\ \citenamefont {Borschevsky}}]{Hao2020}%
  \BibitemOpen
  \bibfield  {author} {\bibinfo {author} {\bibfnamefont {Y.}~\bibnamefont {Hao}}, \bibinfo {author} {\bibfnamefont {P.}~\bibnamefont {Navr\'atil}}, \bibinfo {author} {\bibfnamefont {E.~B.}\ \bibnamefont {Norrgard}}, \bibinfo {author} {\bibfnamefont {M.}~\bibnamefont {Ilia\ifmmode~\check{s}\else \v{s}\fi{}}}, \bibinfo {author} {\bibfnamefont {E.}~\bibnamefont {Eliav}}, \bibinfo {author} {\bibfnamefont {R.~G.~E.}\ \bibnamefont {Timmermans}}, \bibinfo {author} {\bibfnamefont {V.~V.}\ \bibnamefont {Flambaum}},\ and\ \bibinfo {author} {\bibfnamefont {A.}~\bibnamefont {Borschevsky}},\ }\bibfield  {title} {\bibinfo {title} {Nuclear spin-dependent parity-violating effects in light polyatomic molecules},\ }\href {https://doi.org/10.1103/PhysRevA.102.052828} {\bibfield  {journal} {\bibinfo  {journal} {Phys. Rev. A}\ }\textbf {\bibinfo {volume} {102}},\ \bibinfo {pages} {052828} (\bibinfo {year} {2020})}\BibitemShut {NoStop}%
\bibitem [{\citenamefont {Hutzler}(2020)}]{Hutzler2020}%
  \BibitemOpen
  \bibfield  {author} {\bibinfo {author} {\bibfnamefont {N.~R.}\ \bibnamefont {Hutzler}},\ }\bibfield  {title} {\bibinfo {title} {Polyatomic molecules as quantum sensors for fundamental physics},\ }\href {https://doi.org/10.1088/2058-9565/abb9c5} {\bibfield  {journal} {\bibinfo  {journal} {Quantum Science and Technology}\ }\textbf {\bibinfo {volume} {5}},\ \bibinfo {pages} {044011} (\bibinfo {year} {2020})}\BibitemShut {NoStop}%
\bibitem [{\citenamefont {Jadbabaie}\ \emph {et~al.}(2023)\citenamefont {Jadbabaie}, \citenamefont {Takahashi}, \citenamefont {Pilgram}, \citenamefont {Conn}, \citenamefont {Zeng}, \citenamefont {Zhang},\ and\ \citenamefont {Hutzler}}]{Jadbabaie2023}%
  \BibitemOpen
  \bibfield  {author} {\bibinfo {author} {\bibfnamefont {A.}~\bibnamefont {Jadbabaie}}, \bibinfo {author} {\bibfnamefont {Y.}~\bibnamefont {Takahashi}}, \bibinfo {author} {\bibfnamefont {N.~H.}\ \bibnamefont {Pilgram}}, \bibinfo {author} {\bibfnamefont {C.~J.}\ \bibnamefont {Conn}}, \bibinfo {author} {\bibfnamefont {Y.}~\bibnamefont {Zeng}}, \bibinfo {author} {\bibfnamefont {C.}~\bibnamefont {Zhang}},\ and\ \bibinfo {author} {\bibfnamefont {N.~R.}\ \bibnamefont {Hutzler}},\ }\bibfield  {title} {\bibinfo {title} {Characterizing the fundamental bending vibration of a linear polyatomic molecule for symmetry violation searches},\ }\href {https://doi.org/10.1088/1367-2630/ace471} {\bibfield  {journal} {\bibinfo  {journal} {New Journal of Physics}\ }\textbf {\bibinfo {volume} {25}},\ \bibinfo {pages} {073014} (\bibinfo {year} {2023})}\BibitemShut {NoStop}%
\bibitem [{\citenamefont {Touwen}\ \emph {et~al.}(2024)\citenamefont {Touwen}, \citenamefont {van Hofslot}, \citenamefont {Qualm}, \citenamefont {Borchers}, \citenamefont {Bause}, \citenamefont {Bethlem}, \citenamefont {Boeschoten}, \citenamefont {Borschevsky}, \citenamefont {Fikkers}, \citenamefont {Hoekstra}, \citenamefont {Jungmann}, \citenamefont {Marshall}, \citenamefont {Meijknecht}, \citenamefont {Mooij}, \citenamefont {E~Timmermans}, \citenamefont {Ubachs}, \citenamefont {Willmann},\ and\ \citenamefont {eEDM collaboration}}]{Touwen2024}%
  \BibitemOpen
  \bibfield  {author} {\bibinfo {author} {\bibfnamefont {A.}~\bibnamefont {Touwen}}, \bibinfo {author} {\bibfnamefont {J.~W.~F.}\ \bibnamefont {van Hofslot}}, \bibinfo {author} {\bibfnamefont {T.}~\bibnamefont {Qualm}}, \bibinfo {author} {\bibfnamefont {R.}~\bibnamefont {Borchers}}, \bibinfo {author} {\bibfnamefont {R.}~\bibnamefont {Bause}}, \bibinfo {author} {\bibfnamefont {H.~L.}\ \bibnamefont {Bethlem}}, \bibinfo {author} {\bibfnamefont {A.}~\bibnamefont {Boeschoten}}, \bibinfo {author} {\bibfnamefont {A.}~\bibnamefont {Borschevsky}}, \bibinfo {author} {\bibfnamefont {T.~H.}\ \bibnamefont {Fikkers}}, \bibinfo {author} {\bibfnamefont {S.}~\bibnamefont {Hoekstra}}, \bibinfo {author} {\bibfnamefont {K.}~\bibnamefont {Jungmann}}, \bibinfo {author} {\bibfnamefont {V.~R.}\ \bibnamefont {Marshall}}, \bibinfo {author} {\bibfnamefont {T.~B.}\ \bibnamefont {Meijknecht}}, \bibinfo {author} {\bibfnamefont {M.~C.}\ \bibnamefont {Mooij}}, \bibinfo {author} {\bibfnamefont {R.~G.}\ \bibnamefont {E~Timmermans}}, \bibinfo
  {author} {\bibfnamefont {W.}~\bibnamefont {Ubachs}}, \bibinfo {author} {\bibfnamefont {L.}~\bibnamefont {Willmann}},\ and\ \bibinfo {author} {\bibfnamefont {N.}~\bibnamefont {eEDM collaboration}},\ }\bibfield  {title} {\bibinfo {title} {Manipulating a beam of barium fluoride molecules using an electrostatic hexapole},\ }\href {https://doi.org/10.1088/1367-2630/ad60ee} {\bibfield  {journal} {\bibinfo  {journal} {New Journal of Physics}\ }\textbf {\bibinfo {volume} {26}},\ \bibinfo {pages} {073054} (\bibinfo {year} {2024})}\BibitemShut {NoStop}%
\bibitem [{\citenamefont {Flambaum}\ and\ \citenamefont {Murray}(1997)}]{Flambaum1997}%
  \BibitemOpen
  \bibfield  {author} {\bibinfo {author} {\bibfnamefont {V.~V.}\ \bibnamefont {Flambaum}}\ and\ \bibinfo {author} {\bibfnamefont {D.~W.}\ \bibnamefont {Murray}},\ }\bibfield  {title} {\bibinfo {title} {Anapole moment and nucleon weak interactions},\ }\href {https://doi.org/10.1103/PhysRevC.56.1641} {\bibfield  {journal} {\bibinfo  {journal} {Phys. Rev. C}\ }\textbf {\bibinfo {volume} {56}},\ \bibinfo {pages} {1641} (\bibinfo {year} {1997})}\BibitemShut {NoStop}%
\bibitem [{\citenamefont {Collopy}\ \emph {et~al.}(2018)\citenamefont {Collopy}, \citenamefont {Ding}, \citenamefont {Wu}, \citenamefont {Finneran}, \citenamefont {Anderegg}, \citenamefont {Augenbraun}, \citenamefont {Doyle},\ and\ \citenamefont {Ye}}]{Collopy2018}%
  \BibitemOpen
  \bibfield  {author} {\bibinfo {author} {\bibfnamefont {A.~L.}\ \bibnamefont {Collopy}}, \bibinfo {author} {\bibfnamefont {S.}~\bibnamefont {Ding}}, \bibinfo {author} {\bibfnamefont {Y.}~\bibnamefont {Wu}}, \bibinfo {author} {\bibfnamefont {I.~A.}\ \bibnamefont {Finneran}}, \bibinfo {author} {\bibfnamefont {L.}~\bibnamefont {Anderegg}}, \bibinfo {author} {\bibfnamefont {B.~L.}\ \bibnamefont {Augenbraun}}, \bibinfo {author} {\bibfnamefont {J.~M.}\ \bibnamefont {Doyle}},\ and\ \bibinfo {author} {\bibfnamefont {J.}~\bibnamefont {Ye}},\ }\bibfield  {title} {\bibinfo {title} {{3D Magneto-Optical Trap of Yttrium Monoxide}},\ }\href {https://doi.org/10.1103/PhysRevLett.121.213201} {\bibfield  {journal} {\bibinfo  {journal} {Phys. Rev. Lett.}\ }\textbf {\bibinfo {volume} {121}},\ \bibinfo {pages} {213201} (\bibinfo {year} {2018})}\BibitemShut {NoStop}%
\bibitem [{\citenamefont {Höhle}\ \emph {et~al.}(1976)\citenamefont {Höhle}, \citenamefont {Hühnermann}, \citenamefont {Meier}, \citenamefont {Ihle},\ and\ \citenamefont {Wagner}}]{Hoehle1976}%
  \BibitemOpen
  \bibfield  {author} {\bibinfo {author} {\bibfnamefont {C.}~\bibnamefont {Höhle}}, \bibinfo {author} {\bibfnamefont {H.}~\bibnamefont {Hühnermann}}, \bibinfo {author} {\bibfnamefont {T.}~\bibnamefont {Meier}}, \bibinfo {author} {\bibfnamefont {H.}~\bibnamefont {Ihle}},\ and\ \bibinfo {author} {\bibfnamefont {R.}~\bibnamefont {Wagner}},\ }\bibfield  {title} {\bibinfo {title} {Nuclear moments and optical isotope shift of radioactive {133Ba}},\ }\href {https://doi.org/https://doi.org/10.1016/0370-2693(76)90664-X} {\bibfield  {journal} {\bibinfo  {journal} {Physics Letters B}\ }\textbf {\bibinfo {volume} {62}},\ \bibinfo {pages} {390} (\bibinfo {year} {1976})}\BibitemShut {NoStop}%
\bibitem [{\citenamefont {Denis}\ \emph {et~al.}(2020)\citenamefont {Denis}, \citenamefont {Hao}, \citenamefont {Eliav}, \citenamefont {Hutzler}, \citenamefont {Nayak}, \citenamefont {Timmermans},\ and\ \citenamefont {Borschesvky}}]{Denis2020}%
  \BibitemOpen
  \bibfield  {author} {\bibinfo {author} {\bibfnamefont {M.}~\bibnamefont {Denis}}, \bibinfo {author} {\bibfnamefont {Y.}~\bibnamefont {Hao}}, \bibinfo {author} {\bibfnamefont {E.}~\bibnamefont {Eliav}}, \bibinfo {author} {\bibfnamefont {N.~R.}\ \bibnamefont {Hutzler}}, \bibinfo {author} {\bibfnamefont {M.~K.}\ \bibnamefont {Nayak}}, \bibinfo {author} {\bibfnamefont {R.~G.~E.}\ \bibnamefont {Timmermans}},\ and\ \bibinfo {author} {\bibfnamefont {A.}~\bibnamefont {Borschesvky}},\ }\bibfield  {title} {\bibinfo {title} {Enhanced {P,T}-violating nuclear magnetic quadrupole moment effects in laser-coolable molecules},\ }\href {https://doi.org/10.1063/1.5141065} {\bibfield  {journal} {\bibinfo  {journal} {The Journal of Chemical Physics}\ }\textbf {\bibinfo {volume} {152}},\ \bibinfo {pages} {084303} (\bibinfo {year} {2020})}\BibitemShut {NoStop}%
\bibitem [{\citenamefont {Zeng}\ \emph {et~al.}(2023)\citenamefont {Zeng}, \citenamefont {Jadbabaie}, \citenamefont {Patel}, \citenamefont {Yu}, \citenamefont {Steimle},\ and\ \citenamefont {Hutzler}}]{Zeng2023}%
  \BibitemOpen
  \bibfield  {author} {\bibinfo {author} {\bibfnamefont {Y.}~\bibnamefont {Zeng}}, \bibinfo {author} {\bibfnamefont {A.}~\bibnamefont {Jadbabaie}}, \bibinfo {author} {\bibfnamefont {A.~N.}\ \bibnamefont {Patel}}, \bibinfo {author} {\bibfnamefont {P.}~\bibnamefont {Yu}}, \bibinfo {author} {\bibfnamefont {T.~C.}\ \bibnamefont {Steimle}},\ and\ \bibinfo {author} {\bibfnamefont {N.~R.}\ \bibnamefont {Hutzler}},\ }\bibfield  {title} {\bibinfo {title} {Optical cycling in polyatomic molecules with complex hyperfine structure},\ }\href {https://doi.org/10.1103/PhysRevA.108.012813} {\bibfield  {journal} {\bibinfo  {journal} {Phys. Rev. A}\ }\textbf {\bibinfo {volume} {108}},\ \bibinfo {pages} {012813} (\bibinfo {year} {2023})}\BibitemShut {NoStop}%
\bibitem [{\citenamefont {Udrescu}\ \emph {et~al.}(2021)\citenamefont {Udrescu}, \citenamefont {Brinson}, \citenamefont {Ruiz}, \citenamefont {Gaul}, \citenamefont {Berger}, \citenamefont {Billowes}, \citenamefont {Binnersley}, \citenamefont {Bissell}, \citenamefont {Breier}, \citenamefont {Chrysalidis}, \citenamefont {Cocolios}, \citenamefont {Cooper}, \citenamefont {Flanagan}, \citenamefont {Giesen}, \citenamefont {de~Groote}, \citenamefont {Franchoo}, \citenamefont {Gustafsson}, \citenamefont {Isaev}, \citenamefont {Koszor\'us}, \citenamefont {Neyens}, \citenamefont {Perrett}, \citenamefont {Ricketts}, \citenamefont {Rothe}, \citenamefont {Vernon}, \citenamefont {Wendt}, \citenamefont {Wienholtz}, \citenamefont {Wilkins},\ and\ \citenamefont {Yang}}]{Udrescu2021}%
  \BibitemOpen
  \bibfield  {author} {\bibinfo {author} {\bibfnamefont {S.~M.}\ \bibnamefont {Udrescu}}, \bibinfo {author} {\bibfnamefont {A.~J.}\ \bibnamefont {Brinson}}, \bibinfo {author} {\bibfnamefont {R.~F.~G.}\ \bibnamefont {Ruiz}}, \bibinfo {author} {\bibfnamefont {K.}~\bibnamefont {Gaul}}, \bibinfo {author} {\bibfnamefont {R.}~\bibnamefont {Berger}}, \bibinfo {author} {\bibfnamefont {J.}~\bibnamefont {Billowes}}, \bibinfo {author} {\bibfnamefont {C.~L.}\ \bibnamefont {Binnersley}}, \bibinfo {author} {\bibfnamefont {M.~L.}\ \bibnamefont {Bissell}}, \bibinfo {author} {\bibfnamefont {A.~A.}\ \bibnamefont {Breier}}, \bibinfo {author} {\bibfnamefont {K.}~\bibnamefont {Chrysalidis}}, \bibinfo {author} {\bibfnamefont {T.~E.}\ \bibnamefont {Cocolios}}, \bibinfo {author} {\bibfnamefont {B.~S.}\ \bibnamefont {Cooper}}, \bibinfo {author} {\bibfnamefont {K.~T.}\ \bibnamefont {Flanagan}}, \bibinfo {author} {\bibfnamefont {T.~F.}\ \bibnamefont {Giesen}}, \bibinfo {author} {\bibfnamefont {R.~P.}\ \bibnamefont {de~Groote}},
  \bibinfo {author} {\bibfnamefont {S.}~\bibnamefont {Franchoo}}, \bibinfo {author} {\bibfnamefont {F.~P.}\ \bibnamefont {Gustafsson}}, \bibinfo {author} {\bibfnamefont {T.~A.}\ \bibnamefont {Isaev}}, \bibinfo {author} {\bibfnamefont {A.}~\bibnamefont {Koszor\'us}}, \bibinfo {author} {\bibfnamefont {G.}~\bibnamefont {Neyens}}, \bibinfo {author} {\bibfnamefont {H.~A.}\ \bibnamefont {Perrett}}, \bibinfo {author} {\bibfnamefont {C.~M.}\ \bibnamefont {Ricketts}}, \bibinfo {author} {\bibfnamefont {S.}~\bibnamefont {Rothe}}, \bibinfo {author} {\bibfnamefont {A.~R.}\ \bibnamefont {Vernon}}, \bibinfo {author} {\bibfnamefont {K.~D.~A.}\ \bibnamefont {Wendt}}, \bibinfo {author} {\bibfnamefont {F.}~\bibnamefont {Wienholtz}}, \bibinfo {author} {\bibfnamefont {S.~G.}\ \bibnamefont {Wilkins}},\ and\ \bibinfo {author} {\bibfnamefont {X.~F.}\ \bibnamefont {Yang}},\ }\bibfield  {title} {\bibinfo {title} {Isotope shifts of radium monofluoride molecules},\ }\href {https://doi.org/10.1103/PhysRevLett.127.033001} {\bibfield
  {journal} {\bibinfo  {journal} {Phys. Rev. Lett.}\ }\textbf {\bibinfo {volume} {127}},\ \bibinfo {pages} {033001} (\bibinfo {year} {2021})}\BibitemShut {NoStop}%
\bibitem [{\citenamefont {Haxton}\ and\ \citenamefont {Holstein}(2013)}]{Haxton2013}%
  \BibitemOpen
  \bibfield  {author} {\bibinfo {author} {\bibfnamefont {W.~C.}\ \bibnamefont {Haxton}}\ and\ \bibinfo {author} {\bibfnamefont {B.~R.}\ \bibnamefont {Holstein}},\ }\bibfield  {title} {\bibinfo {title} {{Hadronic parity violation}},\ }\href {https://doi.org/https://doi.org/10.1016/j.ppnp.2013.03.009} {\bibfield  {journal} {\bibinfo  {journal} {Progress in Particle and Nuclear Physics}\ }\textbf {\bibinfo {volume} {71}},\ \bibinfo {pages} {185} (\bibinfo {year} {2013})}\BibitemShut {NoStop}%
\end{thebibliography}%

\end{document}